\documentclass[notitlepage,nofootinbib,preprintnumbers,amssymb,superscriptaddress]{revtex4-1}
\usepackage{amsfonts,amssymb,mathtools,graphicx,xcolor,bm}
\definecolor{ultramarine}{rgb}{0.07, 0.04, 0.56}
\definecolor{cadmiumgreen}{rgb}{0.0, 0.42, 0.24}
\definecolor{indigo(dye)}{rgb}{0.0, 0.25, 0.42}
\usepackage[linktocpage=true]{hyperref}
\hypersetup{
colorlinks=true,
citecolor=ultramarine,
linkcolor=cadmiumgreen,
urlcolor=indigo(dye),
}

\usepackage{autobreak}
\usepackage{tikz}
\usepackage{here}
\newcommand{\D}{{\rm d}}
\newcommand{\Ds}{{\rm D}}
\newcommand{\fr}[2]{\frac{#1}{#2}}
\newcommand{\pa}{\partial}
\newcommand{\ti}{\tilde}
\newcommand{\na}{\nabla}
\newcommand{\bra}[1]{\left( #1 \right)}
\newcommand{\brb}[1]{\left[ #1 \right]}

\newcommand{\be}{\begin{equation}}  
\newcommand{\ee}{\end{equation}}
\newcommand{\bem}{\begin{bmatrix}}
\newcommand{\eem}{\end{bmatrix}}

\newcommand{\ga}{\gamma}

\newcommand{\vae}{\varepsilon}

\newcommand{\la}{\lambda}

\newcommand{\mn}{{\mu \nu}}
\newcommand{\mB}{\mathcal{B}}
\newcommand{\mC}{\mathcal{C}}
\newcommand{\mD}{\mathcal{D}}
\newcommand{\mF}{\mathcal{F}}
\newcommand{\mG}{\mathcal{G}}
\newcommand{\mJ}{\mathcal{J}}
\newcommand{\mK}{\mathcal{K}}
\newcommand{\mL}{\mathcal{L}}
\newcommand{\mM}{\mathcal{M}}
\newcommand{\mN}{\mathcal{N}}
\newcommand{\mW}{\mathcal{W}}
\newcommand{\kinmat}{\bar{X}_\sigma}
\newcommand{\mFc}{f}
\newcommand{\sbt}{~\begin{picture}(-1,1)(-1,-2.5)\circle*{3}\end{picture}~\;}

\begin{document}

\preprint{YITP-22-89}

\title{Generalized disformal Horndeski theories: \texorpdfstring{\\}{}
cosmological perturbations and consistent matter coupling}

\author{Kazufumi Takahashi}
\affiliation{Center for Gravitational Physics and Quantum Information, Yukawa Institute for Theoretical Physics, Kyoto University, 606-8502, Kyoto, Japan}

\author{Masato Minamitsuji}
\affiliation{Centro de Astrof\'{\i}sica e Gravita\c c\~ao  - CENTRA, Departamento de F\'{\i}sica, Instituto Superior T\'ecnico - IST,
Universidade de Lisboa - UL, Av.~Rovisco Pais 1, 1049-001 Lisboa, Portugal}

\author{Hayato Motohashi}
\affiliation{Division of Liberal Arts, Kogakuin University, 2665-1 Nakano-machi, Hachioji, Tokyo 192-0015, Japan}

\begin{abstract}
Invertible disformal transformations are a useful tool to investigate ghost-free scalar-tensor theories.
By performing a higher-derivative generalization of the invertible disformal transformation on Horndeski theories, we construct a novel class of ghost-free scalar-tensor theories, which we dub {\it generalized disformal Horndeski theories}.
Specifically, these theories lie beyond the quadratic/cubic DHOST class.
We explore cosmological perturbations to identify a subclass where gravitational waves propagate at the speed of light and clarify the conditions for the absence of ghost/gradient instabilities for tensor and scalar perturbations.
We also investigate the conditions under which a matter field can be consistently coupled to these theories without introducing unwanted extra degrees of freedom.
\end{abstract}

\maketitle

\section{Introduction}\label{sec:intro}

Scalar-tensor theories are a simplest and robust framework of modified gravity and they have been employed as a model of early/late-time universe as well as black holes/neutron stars.
In particular, there have been extensive studies on theories without the problem of Ostrogradsky ghosts~\cite{Woodard:2015zca,Motohashi:2020psc}, i.e., unstable extra degrees of freedom (DOFs) associated with nondegenerate higher derivatives in the equations of motion.
The trend to search for such ghost-free scalar-tensor theories has accelerated after the rediscovery of Horndeski theories~\cite{Horndeski:1974wa} (i.e., the most general scalar-tensor theories with second-order Euler-Lagrange equations) in the context of generalized Galileons~\cite{Deffayet:2011gz,Kobayashi:2011nu}.
A conventional way to construct ghost-free scalar-tensor theories is to specify the form of the Lagrangian and then impose the so-called degeneracy conditions to avoid the problem of Ostrogradsky ghosts~\cite{Motohashi:2014opa,Langlois:2015cwa,Motohashi:2016ftl,Klein:2016aiq,Motohashi:2017eya,Motohashi:2018pxg}.
For instance, performing this program for the Lagrangian that contains the second derivative of the scalar field up to the quadratic or cubic order (supplemented with appropriate interactions with the curvature tensor), one obtains the so-called quadratic/cubic degenerate higher-order scalar-tensor (DHOST) theories~\cite{Langlois:2015cwa,Crisostomi:2016czh,BenAchour:2016fzp}.\footnote{One could require the degeneracy only in the unitary gauge, in which case one obtains a broader framework called U-DHOST theories~\cite{DeFelice:2018mkq,DeFelice:2021hps,DeFelice:2022xvq}.
In U-DHOST theories, there is an apparent Ostrogradsky mode, but it satisfies an elliptic differential equation on a spacelike hypersurface and hence does not propagate.}
(See \cite{Langlois:2018dxi,Kobayashi:2019hrl} for reviews.)

Another systematic way to obtain a ghost-free scalar-tensor theory is to perform an invertible transformation on known ghost-free scalar-tensor theories~\cite{Zumalacarregui:2013pma}.
In the context of scalar-tensor theories, one can consider the so-called disformal transformation~\cite{Bekenstein:1992pj,Bruneton:2007si,Bettoni:2013diz}, which defines a map from the space of scalar-tensor theories to itself.
A disformal transformation maps a pair of the metric~$g_\mn$ and the scalar field~$\phi$ to another one~$(\bar{g}_\mn,\phi)$ with the scalar field unchanged, where
    \be
    \bar{g}_\mn[g,\phi] = F_0(\phi,X) g_\mn + F_1(\phi,X)\na_\mu\phi\na_\nu\phi.
    \label{disformal1_intro}
    \ee
Here, $F_0$ and $F_1$ are functions of $(\phi,X)$ with $X\coloneqq g^{\alpha\beta}\na_\alpha\phi\na_\beta\phi$.
Clearly, this is a generalization of the conformal transformation for which $F_1=0$.
Note that the right-hand side of \eqref{disformal1_intro} is the most general symmetric rank-two tensor constructed from the metric, the scalar field, and its first derivative.
The above transformation is known to be invertible, i.e., $g_\mn$ can be uniquely expressed as a functional of $\bar{g}_\mn$ and $\phi$ at least locally, so long as the following conditions are satisfied~\cite{Zumalacarregui:2013pma,Takahashi:2021ttd}:
    \be
    F_0 \ne 0, \qquad F_0+XF_1 \ne 0, \qquad F_0-XF_{0X}-X^2F_{1X} \ne 0, \label{inv_cond1_intro}
    \ee
with a subscript~$X$ denoting the partial derivative with respect to $X$.
For a given action~$S[g_\mn,\phi]$, the replacement~$g_\mn\to \bar{g}_\mn[g,\phi]$ yields a new action~$\ti{S}[g_\mn,\phi]\coloneqq S[\bar{g}_\mn,\phi]$.
It is known that an invertible transformation does not change the number of physical DOFs, and hence the absence of Ostrogradsky ghosts should be preserved under invertible transformations~\cite{Domenech:2015tca,Takahashi:2017zgr}.
If one performs the disformal transformation~\eqref{disformal1_intro} on Horndeski theories, one obtains a subclass of the quadratic/cubic DHOST class, which we shall dub {\it the disformal Horndeski class}.
In terms of the classification in \cite{Crisostomi:2016czh,BenAchour:2016fzp}, this is a sum of the quadratic DHOST of class ${}^2$N-I (also called class Ia in \cite{Achour:2016rkg}) and the cubic DHOST of class ${}^3$N-I.\footnote{Precisely speaking, one cannot freely add the Lagrangian of quadratic DHOST and that of cubic DHOST because the degeneracy structure of the former is incompatible with that of the latter in general.
In order for the Lagrangian to be degenerate as a whole, one has to require additional degeneracy conditions which relate coefficient functions in the quadratic sector and those in the cubic sector~\cite{BenAchour:2016fzp}.
A similar problem appears when matter field(s) are coupled to DHOST theories as we discuss later.}
Of course, one could further perform the disformal transformation~\eqref{disformal1_intro} on quadratic/cubic DHOST theories, but this does not yield a new class of theories.
Therefore, the disformal transformation of the form~\eqref{disformal1_intro} is not helpful for further enlarging the class of ghost-free scalar-tensor theories.
Interestingly, from a phenomenologically point of view, disformal Horndeski theories provide the only viable class within quadratic/cubic DHOST theories:
Those lying outside the disformal Horndeski class are disfavored because they do not accommodate stable cosmological solution or otherwise the tensor perturbations are nondynamical~\cite{Langlois:2017mxy}.
Due to this reason, the term ``quadratic/cubic DHOST theories'' is often used to mean disformal Horndeski theories in the literature.

When two actions of scalar-tensor theories are related by an invertible disformal transformation, they are mathematically equivalent (up to boundary terms).
Hence, one might think that theories generated by an invertible transformation of known ones are not essentially new.
However, once matter fields are taken into account, the two theories are no longer related by the invertible transformation in general, and hence they can be distinguished from each other.
Related to this issue, the introduction of matter fields could introduce extra DOFs,
namely, the number of DOFs of a matter-coupled scalar-tensor theory can be different from the sum of the numbers of DOFs of the gravitational and matter sectors~\cite{Deffayet:2020ypa}.
This happens because the matter sector does not respect the degeneracy conditions imposed on the gravitational sector in general.
As such, the extra DOFs would be Ostrogradsky ghosts, and hence one would like to avoid them.

Along this line of thought, the authors of the present paper developed a higher-derivative generalization of the disformal transformation, in which the second (covariant) derivative of the scalar field shows up~\cite{Takahashi:2021ttd}.
We focused on the group structure under the functional composition of the generalized disformal transformation to clarify the conditions under which the transformation is invertible (see \cite{Takahashi:2021ttd} or \S\ref{ssec:higher} below for details).
In the present paper, we perform this invertible generalized disformal transformation on Horndeski theories to generate a novel class of ghost-free scalar-tensor theories, which we dub {\it generalized disformal Horndeski theories}.
By construction, generalized disformal Horndeski theories include disformal Horndeski theories, i.e., the phenomenologically viable subclass of quadratic/cubic DHOST theories, and more general ghost-free theories beyond them.
We then specify the subclass of generalized disformal Horndeski theories where gravitational waves propagate at the speed of light, which is consistent with the almost simultaneous detection of the gravitational-wave event~GW170817 and the gamma-ray burst~170817A emitted from a binary neutron star merger~\cite{TheLIGOScientific:2017qsa,GBM:2017lvd,Monitor:2017mdv}.
Moreover, we study the condition under which a matter field can be consistently coupled without introducing unwanted extra DOFs.

The rest of this paper is organized as follows.
In \S\ref{sec:inv}, we review the construction of invertible generalized disformal transformations based on \cite{Takahashi:2021ttd}.
We then perform the transformation on Horndeski theories to write down the action of generalized disformal Horndeski theories.
In \S\ref{sec:cosmology}, we study homogeneous and isotropic cosmology in generalized disformal Horndeski theories.
In \S\ref{sec:coupling}, we investigate the consistency of matter coupling in generalized disformal Horndeski theories.
Finally, we draw our conclusions in \S\ref{sec:conc}.

\section{Generalized disformal transformations}\label{sec:inv}

\subsection{Invertible disformal transformations with higher derivatives}\label{ssec:higher}

In this subsection, we briefly review the invertible disformal transformation with second derivatives of the scalar field constructed in our previous study~\cite{Takahashi:2021ttd}.
We consider a generalized disformal transformation defined by
    \be
    \bar{g}_\mn[g,\phi] = F_0 g_\mn + F_1 \phi_\mu\phi_\nu
    + 2F_2 \phi_{(\mu}X_{\nu)} + F_3 X_\mu X_\nu, \label{disformal2}
    \ee
where $\phi_\mu\coloneqq \partial_\mu\phi$ and $X_\mu\coloneqq \pa_\mu X=2\phi_\alpha\phi^\alpha_\mu$ with $\phi_\mn\coloneqq \na_\mu\na_\nu\phi$ and $\phi^\alpha_\mu=g^{\alpha\beta}\phi_{\beta\mu}$.
Also, we have defined the symmetrization of a tensor~$T_\mn$ by $T_{(\mn)}\coloneqq (T_\mn+T_{\nu\mu})/2$.
Here, $F_i$'s ($i=0,1,2,3$) are functions of $(\phi,X,Y,Z)$, with $Y$ and $Z$ defined as
    \be
    Y\coloneqq \phi_\mu X^\mu, \qquad
    Z\coloneqq X_\mu X^\mu. \label{YandZ}
    \ee
We define the following quantities for later use:
    \be
    \mF\coloneqq F_0^2+F_0(XF_1+2YF_2+ZF_3)+W(F_2^2-F_1F_3), \qquad
    W\coloneqq Y^2-XZ.
    \label{mF}
    \ee
Note that the class of conventional disformal transformations of the form~\eqref{disformal1_intro} is included in the class of generalized disformal transformations given by \eqref{disformal2}.

Let $\mD$ denote the set of disformal transformations of the form~\eqref{disformal2}.
In order to discuss the inverse transformation, we consider a subset of $\mD$ that forms groups under the following two binary operations:
\begin{itemize}
\item {\it Matrix product}:\footnote{Strictly speaking, in order to guarantee the closedness under the matrix product, we need to enlarge the underlying set~$\mD$ to include the antisymmetric part of $\phi_\mu X_\nu$~\cite{Takahashi:2021ttd}. Nevertheless, the inverse element of an (invertible) element in $\mD$ can also be found in $\mD$, which allows us to construct the inverse disformal metric~\eqref{inv_met2}.}
    \be
    \mD^2\ni (\bar{g}_\mn,\hat{g}_\mn)\quad \mapsto \quad
    (\bar{g}\sbt\hat{g})_\mn\coloneqq
    \bar{g}_{\mu\alpha}g^{\alpha\beta}\hat{g}_{\beta\nu}. \label{matrix_product}
    \ee
\item {\it Functional composition}:
    \be
    \mD^2\ni (\bar{g}_\mn[g,\phi],\hat{g}_\mn[g,\phi])\quad \mapsto \quad
    (\bar{g}\circ\hat{g})_\mn\coloneqq
    \bar{g}_\mn[\hat{g},\phi]. \label{functional_composition}
    \ee
\end{itemize}
Note that both the operations are associative and have an identity element (given by $g_\mn$ itself).
The inverse element of $\bar{g}_\mn$ associated with the matrix product (after raising the indices by $g^{\alpha\beta}$) gives the inverse metric~$\bar{g}^\mn$, while the one associated with the functional composition is nothing but the inverse transformation~$g_\mn[\bar{g},\phi]$.
We shall clarify which subset of $\mD$ can form groups under these operations.

Provided that $F_0\ne 0$ and $\mF\ne 0$, it is straightforward to obtain the inverse metric with respect to $\bar{g}_\mn$~\cite{Takahashi:2021ttd}:
    \be
    \bar{g}^\mn=\fr{1}{F_0}\bigg[g^\mn-\fr{F_0F_1-Z(F_2^2-F_1F_3)}{\mF}\phi^\mu\phi^\nu
    -2\fr{F_0F_2+Y(F_2^2-F_1F_3)}{\mF}\phi^{(\mu}X^{\nu)}-\fr{F_0F_3-X(F_2^2-F_1F_3)}{\mF}X^\mu X^\nu\bigg]. \label{inv_met2}
    \ee
With the inverse metric, the kinetic term in the barred frame can be computed as
    \be
    \bar{X}
    =\bar{g}^\mn \phi_\mu\phi_\nu
    =\fr{XF_0-WF_3}{\mF}, 
    \label{Xbar2}
    \ee
which is a function of $(\phi,X,Y,Z)$ in general.
If $\bar{X}$ depends on $Y$ and/or $Z$ in a nontrivial manner, the quantities~$\bar{Y}$ and $\bar{Z}$, which are respectively the barred counterparts of $Y$ and $Z$ defined in \eqref{YandZ}, yield unwanted higher derivatives like $\pa_\mu Y$ and $\pa_\mu Z$ through $\pa_\mu \bar{X}$.
Therefore, we require
    \be
    \bar{X}_Y=\bar{X}_Z=0, \label{Xbar_cond}
    \ee
where 
${\bar X}_Y\coloneqq \partial {\bar X}/\partial Y$
and ${\bar X}_Z\coloneqq \partial {\bar X}/\partial Z$, so that $\bar{X}=\bar{X}(\phi,X)$.
We also assume $\bar{X}_X\ne 0$ so that the relation~$\bar{X}=\bar{X}(\phi,X)$ can be solved for $X$ to yield $X=X(\phi,\bar{X})$.
Then, we have $\bar{X}_\mu\coloneqq \pa_\mu \bar{X}= \bar{X}_X X_\mu + \bar{X}_\phi \phi_\mu$, and hence
    \be
    \begin{split}
    \bar{Y}&=\bar{g}^\mn\phi_\mu \bar{X}_\nu
    =\bar{X}_X\fr{Y F_0 + W F_2}{\mF}+\bar{X}_\phi\bar{X}, \\
    \bar{Z}&=\bar{g}^\mn\bar{X}_\mu \bar{X}_\nu
    =\bar{X}_X^2\fr{ZF_0-WF_1}{\mF}+2\bar{X}_\phi\bar{Y}-\bar{X}_\phi^2\bar{X}.
    \end{split} \label{YZbar}
    \ee
We further require that these two equations can be solved for $Y$ and $Z$ to obtain $Y=Y(\phi,\bar{X},\bar{Y},\bar{Z})$ and $Z=Z(\phi,\bar{X},\bar{Y},\bar{Z})$, which is possible if the Jacobian determinant~$|\pa(\bar{Y},\bar{Z})/\pa(Y,Z)|$ is nonvanishing.
As a side note, the quantity~$W$ is transformed as
    \be
    \bar{W}\coloneqq \bar{Y}^2-\bar{X}\bar{Z}=\fr{\bar{X}_X^2}{\mF}W. \label{Wbar}
    \ee

The requirements discussed above provide the invertibility condition, namely~\cite{Takahashi:2021ttd},
    \be
    F_0\ne 0, \qquad
    \mF\ne 0, \qquad
    \bar{X}_Y=\bar{X}_Z=0, \qquad
    \bar{X}_X\ne 0, \qquad
    \left|\fr{\pa(\bar{Y},\bar{Z})}{\pa(Y,Z)}\right|\ne 0.
    \label{inv_cond}
    \ee
Under this condition, we can obtain the inverse transformation for \eqref{disformal2} in the following form~\cite{Takahashi:2021ttd}:
    \be
    g_\mn=\fr{1}{F_0}\bra{\bar{g}_\mn -\fr{\bar{X}_X^2F_1-2\bar{X}_\phi\bar{X}_XF_2+\bar{X}_\phi^2F_3}{\bar{X}_X^2}\phi_\mu\phi_\nu
    -2\fr{\bar{X}_XF_2-\bar{X}_\phi F_3}{\bar{X}_X^2}\phi_{(\mu}\bar{X}_{\nu)} -\fr{F_3}{\bar{X}_X^2}\bar{X}_\mu \bar{X}_\nu}, \label{inv_trnsf2}
    \ee
where the functions of $(\phi,X,Y,Z)$ in the right-hand side can be translated back into functions of $(\phi,\bar{X},\bar{Y},\bar{Z})$ by use of \eqref{Xbar2} and \eqref{YZbar}.
In other words, the set of conditions~\eqref{inv_cond} defines a subset of $\mD$ that forms groups under the matrix product~\eqref{matrix_product} and the functional composition~\eqref{functional_composition}.

Thus, we clarified the invertibility condition~\eqref{inv_cond} for the generalized disformal transformation~\eqref{disformal2}.
Before closing this subsection, there are several things to note about the invertible generalized disformal transformation and its possible further extensions.

\begin{enumerate}
\item The first is about the independent functional DOFs of invertible transformations.
As we already mentioned in \cite{Takahashi:2021ttd}, one can regard $\bar{X}=\bar{X}(\phi,X)$ as a given function, which then fixes one of the coefficient functions, say $F_3$, in the generalized disformal transformation~\eqref{disformal2} through \eqref{Xbar2}.
Written explicitly,
    \be
    F_3=\fr{XF_0-\bar{X}(\phi,X)\brb{F_0(F_0+XF_1+2YF_2)+WF_2^2}}{W+\bar{X}(\phi,X)\bra{ZF_0-WF_1}}. \label{F3}
    \ee
Therefore, the invertible subclass of transformations is characterized by $\bar{X}(\phi,X)$ as well as the remaining three functions~$F_0$, $F_1$, $F_2$ of $(\phi,X,Y,Z)$.
Then, the third condition in \eqref{inv_cond} is trivial, and the independent functional DOFs should be chosen to satisfy the remaining four conditions.

\item The invertibility condition~\eqref{inv_cond} for generalized disformal transformations is precisely a generalization of \eqref{inv_cond1_intro} for conventional disformal transformations.
Indeed, by choosing the independent functional DOFs of generalized disformal transformations as
    \be
    F_0=F_0(\phi,X), \qquad
    F_1=F_1(\phi,X), \qquad
    F_2=0, \qquad
    \bar{X}=\fr{X}{F_0+XF_1},
    \ee
we have $F_3=0$ from \eqref{F3}, and hence the transformation law~\eqref{disformal2} reduces to the conventional one~\eqref{disformal1_intro}.
Then, the last two conditions in \eqref{inv_cond} are degenerate, which yield $\bar{X}_X\propto F_0-XF_{0X}-X^2F_{1X} \ne 0$. 
Also, we now have $\mF\propto F_0+XF_1\ne 0$.
Therefore, we recover \eqref{inv_cond1_intro} from \eqref{inv_cond}.

\item The inverse transformation may not be unique in the following sense.
We required the last two conditions in \eqref{inv_cond} so that equations~\eqref{Xbar2} and \eqref{YZbar} can be solved algebraically for $X$, $Y$, and $Z$.
Since these equations are nonlinear algebraic equations in general, there could be multiple branches of (real) solutions.
In this case, we have the inverse transformation for each branch of solution.

\item It could happen that the set of conditions~\eqref{inv_cond} is violated only at some spacetime point(s) for some particular configurations.
If this happens for a solution in some scalar-tensor theory, then the configuration obtained as a result of the disformal transformation cannot be regarded as a solution in the disformal counterpart of the scalar-tensor theory in general.
In other words, such a transformation could introduce a new branch of solution which is not connected to solutions in the original theory~\cite{Jirousek:2022rym,Jirousek:2022jhh}.

\item The transformation law~\eqref{disformal2} is not most general up to the second derivative of the scalar field, and there could be other types of invertible transformations 
(see \cite{Domenech:2019syf} for an example constructed based on conformally/disformally invariant quantities).
In principle, it would be possible to find a more general class of invertible disformal transformations with the second derivative of the scalar field, which we leave for future work.

\item Finally, it is also possible to construct invertible disformal transformations with third or higher derivatives of the scalar field~\cite{Takahashi:2021ttd}.
Moreover, as we shall discuss in Appendix~\ref{AppA}, one can incorporate the curvature tensor in the transformation law.

\end{enumerate}

\subsection{Generalized disformal Horndeski theories}\label{ssec:disformal_Horndeski}

Let us apply the generalized disformal transformation~\eqref{disformal2} satisfying the invertibility condition~\eqref{inv_cond} to generate novel ghost-free theories beyond the quadratic/cubic DHOST class.
For a given action~$S_{\rm g}[g_\mn,\phi]$ of scalar-tensor theories, one can replace the metric by the generalized disformal metric~\eqref{disformal2} to obtain a new action, i.e.,
    \begin{align}
    S_{\rm g}[g_\mn,\phi] \quad \mapsto \quad
    \ti{S}_{\rm g}[g_\mn,\phi]\coloneqq S_{\rm g}[\bar{g}_\mn,\phi].
    \label{disformal_action}
    \end{align}
Suppose that the generalized disformal metric satisfies the invertibility condition~\eqref{inv_cond}.
Since the two gravitational actions~$S_{\rm g}[g_\mn,\phi]$ and $\ti{S}_{\rm g}[g_\mn,\phi]$ are related to each other
by invertible transformation, the two theories are mathematically equivalent in the absence of matter fields.
However, the theory described by the action~$\ti{S}_{\rm g}[g_\mn,\phi]$ should be distinguished from the seed theory when matter fields are taken into account.
Let us assume matter fields (which we denote collectively by $\Psi$) minimally coupled to gravity and consider the following two actions:
    \be
    S[g_\mn,\phi,\Psi]\coloneqq S_{\rm g}[g_\mn,\phi]+S_{\rm m}[g_\mn,\Psi], \qquad
    \ti{S}[g_\mn,\phi,\Psi]\coloneqq \ti{S}_{\rm g}[g_\mn,\phi]+S_{\rm m}[g_\mn,\Psi].
    \label{disf-related_theories}
    \ee
Due to the existence of the matter fields, the two actions are no longer disformally related to each other.
Indeed, performing the disformal transformation~$(g_\mn,\phi)\mapsto (\bar{g}_\mn,\phi)$ on the first action, we obtain
    \be
    S[g_\mn,\phi,\Psi] \quad \mapsto \quad
    S[\bar{g}_\mn,\phi,\Psi]=\ti{S}_{\rm g}[g_\mn,\phi]+S_{\rm m}[\bar{g}_\mn,\Psi],
    \ee
where we have used $S_{\rm g}[\bar{g}_\mn,\phi]=\ti{S}_{\rm g}[g_\mn,\phi]$.
This is different from $\ti{S}[g_\mn,\phi,\Psi]$ because the matter fields are minimally coupled to $\bar{g}_\mn$, not $g_\mn$.
In this sense, the two theories are no longer equivalent in the presence of matter fields.

In \cite{Takahashi:2021ttd}, we discussed the transformation of a general action via the generalized disformal transformation~\eqref{disformal2} and demonstrated that the transformation of a simplest scalar-tensor theory indeed generates a nontrivial higher-derivative theory.
In what follows, we study how Horndeski theories, which form a large class of ghost-free scalar-tensor theories, are transformed under the generalized disformal transformation and write down the resultant action explicitly.
As mentioned earlier, we are interested in the invertible subset of transformations, by which any ghost-free theory is mapped to another ghost-free theory.
Indeed, this is the case when one performs the conventional disformal transformation on Horndeski theories, and we call the resultant class of theories {\it the disformal Horndeski class} (or the DH class for short).
The DH class is (literally) a part of the DHOST class:
It is a sum of the quadratic DHOST of class ${}^2$N-I and the cubic DHOST of class ${}^3$N-I~\cite{Crisostomi:2016czh,BenAchour:2016fzp}.
We will see that the generalized disformal transformation of Horndeski theories in general lies beyond 
the quadratic/cubic DHOST class.
We call the resultant class of novel ghost-free scalar-tensor theories {\it the generalized disformal Horndeski class} (or the GDH class).

The action of Horndeski theories (in four spacetime dimensions) are given by
    \begin{align}
    S_{\rm H}[g_\mn,\phi]
    =\int \D^4x\sqrt{-g}\sum_{I=2}^{5}\mL_I[g_\mn,\phi], \label{Horndeski}
    \end{align}
with
    \be
    \begin{split}
    \mL_2&\coloneqq G_2(\phi,X), \\
    \mL_3&\coloneqq G_3(\phi,X)\Box\phi, \\
    \mL_4&\coloneqq G_4(\phi,X)R-4G_{4X}(\phi,X)g^{\alpha[\mu}g^{\nu]\beta}\phi_{\alpha\mu}\phi_{\beta\nu}, \\
    \mL_5&\coloneqq G_5(\phi,X)G^\mn\phi_\mn+2G_{5X}(\phi,X)g^{\alpha[\mu|}g^{\beta|\nu|}g^{\gamma|\lambda]}\phi_{\alpha\mu}\phi_{\beta\nu}\phi_{\gamma\lambda}.
    \end{split}
    \ee
Here, $G_I$'s are arbitrary functions of $(\phi,X)$, $G_\mn$ denotes the Einstein tensor, and indices inside square brackets are anti-symmetrized (note that $\mu$, $\nu$, and $\lambda$ are anti-symmetrized in the second term of $\mL_5$).
The generalized disformal transformation of the action~\eqref{Horndeski} is obtained by the replacement~$g_\mn\to \bar{g}_\mn$, with $\bar{g}_\mn$ being a functional of $g_\mn$ and $\phi$ given by \eqref{disformal2}.
By employing the transformation law for each building block of the action developed in \cite{Takahashi:2021ttd}, it is straightforward to compute the generalized disformal transformation of the action~\eqref{Horndeski}.
Written explicitly, the action of GDH theories is given by
    \be
    S_{\rm GDH}[g_\mn,\phi]\coloneqq S_{\rm H}[\bar{g}_\mn,\phi]
    =\int \D^4x\sqrt{-g}\,\mJ\sum_{I=2}^{5}\ti{\mL}_I[g_\mn,\phi], \label{GDH}
    \ee
where $\mJ\coloneqq F_0\mF^{1/2}$ and
    \be
    \begin{split}
    \ti{\mL}_2&\coloneqq G_2, \\
    \ti{\mL}_3&\coloneqq G_3\bar{g}^\mn(\phi_\mn-C^\rho{}_\mn\phi_\rho), \\
    \ti{\mL}_4&\coloneqq G_4\bar{g}^\mn\bra{R_{\mn}-2C^\alpha{}_{\beta[\alpha}C^\beta{}_{\nu]\mu}} - 2 \bra{G_{4\phi}\phi_\alpha+G_{4X}\bar{X}_\alpha} \bar{g}^{\mu[\nu} C^{\alpha]}{}_{\mn} \\
    &\quad\: -4G_{4X}\bar{g}^{\alpha[\mu}\bar{g}^{\nu]\beta}(\phi_{\alpha\mu}-C^\sigma{}_{\alpha\mu}\phi_\sigma)(\phi_{\beta\nu}-C^\rho{}_{\beta\nu}\phi_\rho), \\
    \ti{\mL}_5&\coloneqq G_5\bra{\bar{g}^{\mu\lambda}\bar{g}^{\nu\sigma}-\fr{1}{2}\bar{g}^{\mn}\bar{g}^{\lambda\sigma}}
    (\phi_\mn-C^\rho{}_\mn\phi_\rho)
    \bra{R_{\lambda\sigma}+2\na_{[\alpha} C^\alpha{}_{\sigma]\lambda}+2C^\alpha{}_{\beta[\alpha}C^\beta{}_{\sigma]\lambda}} \\
    &\quad\: +2G_{5X}\bar{g}^{\alpha[\mu|}\bar{g}^{\beta|\nu|}\bar{g}^{\gamma|\lambda]}(\phi_{\alpha\mu}-C^\sigma{}_{\alpha\mu}\phi_\sigma)(\phi_{\beta\nu}-C^\rho{}_{\beta\nu}\phi_\rho)(\phi_{\gamma\lambda}-C^\eta{}_{\gamma\lambda}\phi_\eta),
    \end{split}
    \ee
with all $G_I$'s and their derivatives evaluated at $(\phi,\bar{X})$.
Here, we have defined the tensor~$C$ as the change of the Christoffel symbol under the generalized disformal transformation:
    \be
    C^\la{}_\mn\coloneqq \bar{\Gamma}^\la_\mn-\Gamma^\la_\mn
    =\bar{g}^{\la\alpha}\bra{\na_{(\mu}\bar{g}_{\nu)\alpha} -\fr{1}{2}\na_\alpha\bar{g}_\mn}, \label{Ctensor}
    \ee
which contains the third derivative of $\phi$ through $\na\bar{g}$ since $\bar{g}$ itself contains the second derivative of $\phi$ [see Eq.~\eqref{disformal2}].
Note that we have performed an integration by parts to remove a covariant derivative acting on the tensor~$C$ in $\ti{\mL}_4$.
There still remain terms containing $\na C$ and hence the fourth derivatives of $\phi$, but they can also be removed by integration by parts.\footnote{The authors would like to thank Hiroaki W.~H.~Tahara for useful discussion on this point.}
This means that the action~\eqref{GDH} of GDH theories can be recast in a form where derivatives of the scalar field appear only up to the cubic order.

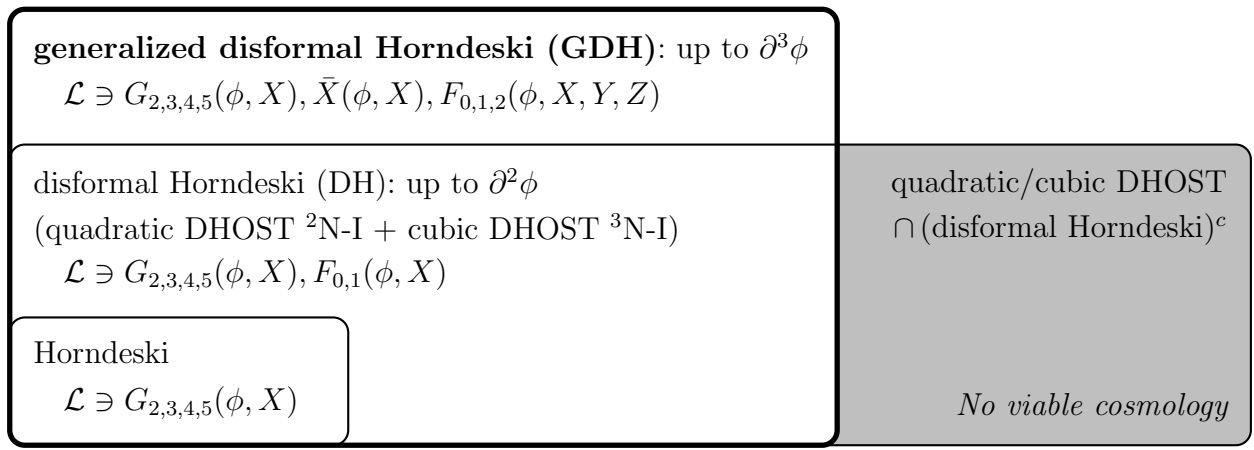
\begin{figure}[H]
\centering
\begin{tikzpicture}
	\fill[lightgray, rounded corners=0.2cm] (0,0)--(0,4)--(16.5,4)--(16.5,0)--cycle;
	\fill[white, rounded corners=0.2cm] (0,0)--(0,6)--(11,6)--(11,0)--cycle;
	\draw[thick, rounded corners=0.2cm] (0,0)--(0,1.7)--(4.5,1.7)--(4.5,0)--cycle;
	\node[anchor=north west] at (0.2,1.5) {\large{Horndeski}};
	\node[anchor=north west] at (0.6,0.9) {\large{$\mL\ni G_{2,3,4,5}(\phi,X)$}};
	\draw[thick, rounded corners=0.2cm] (0,0)--(0,4)--(16.5,4)--(16.5,0)--cycle;
	\node[anchor=north east] at (16.3,3.8) {\large{quadratic/cubic DHOST}};
	\node[anchor=north east] at (16.3,3.2) {\large{$\cap\,(\text{disformal Horndeski})^c$}};
	\node[anchor=south east] at (16.3,0.2) {\large{\it{No 
	viable cosmology}}};
	\node[anchor=north west] at (0.2,3.8) {\large{disformal Horndeski (DH): up to $\pa^2\phi$}};
	\node[anchor=north west] at (0.2,3.2) {\large{(quadratic DHOST ${}^2$N-I $+$ cubic DHOST ${}^3$N-I)}};
	\node[anchor=north west] at (0.6,2.6) {\large{$\mL\ni G_{2,3,4,5}(\phi,X), F_{0,1}(\phi,X)$}};
	\draw[line width=2pt, rounded corners=0.2cm] (0,0)--(0,5.8)--(11,5.8)--(11,0)--cycle;
	\node[anchor=north west] at (0.2,5.6) {\large{{\bf generalized disformal Horndeski (GDH)}: up to $\pa^3\phi$}};
	\node[anchor=north west] at (0.6,5) {\large{$\mL\ni G_{2,3,4,5}(\phi,X), \bar{X}(\phi,X), F_{0,1,2}(\phi,X,Y,Z)$}};
\end{tikzpicture}
\caption{Inclusion relation among ghost-free scalar-tensor theories.
Independent functional DOFs are shown for each theory.
The DH class is a part of the DHOST class. 
It is related to the Horndeski class via the conventional disformal transformation~\eqref{disformal1_intro} and can accommodate viable cosmological solutions.
The GDH class is a generalization of the DH class, which is related to the Horndeski class via the generalized disformal transformation~\eqref{disformal2}.}
\label{fig1}
\end{figure}

In Fig.~\ref{fig1}, we illustrate the inclusion relation among ghost-free scalar-tensor theories.
The GDH theory~\eqref{GDH} is characterized by the functions~$G_I(\phi,X)$ ($I=2,3,4,5$) associated with the seed Horndeski theory and the functions~$\bar{X}(\phi,X)$ and $F_i(\phi,X,Y,Z)$ ($i=0,1,2$) associated with the invertible generalized disformal transformation.
As mentioned above, the GDH class involves known ghost-free scalar-tensor theories as special subclasses. 
Clearly, it includes the Horndeski class as a subclass, which amounts to considering the identity transformation (i.e., $F_0=1$, $F_1=F_2=F_3=0$).
For conventional disformal transformations characterized by $F_0(\phi,X)$ and $F_1(\phi,X)$ satisfying the invertibility condition~\eqref{inv_cond1_intro}, the action~\eqref{GDH} reduces to DH theories, though the entire class of quadratic/cubic DHOST theories is not covered by GDH theories.
It is known that the complement of the DH subclass in the quadratic/cubic DHOST class, which corresponds to the shaded region in Fig.~\ref{fig1}, does not accommodate stable cosmological solution or otherwise the tensor perturbations are nondynamical~\cite{Langlois:2017mxy}.
Hence, from a phenomenologically point of view, DH theories provide the only viable class within quadratic/cubic DHOST theories.
With generic $F_i(\phi,X,Y,Z)$ satisfying the invertibility condition~\eqref{inv_cond}, the generalized disformal transformation allows us to go beyond quadratic/cubic DHOST theories and obtain a wider framework of GDH theories~\eqref{GDH}, which are free from Ostrogradsky ghosts and are expected to admit stable and viable cosmological solutions.

\section{Homogeneous and isotropic cosmology}\label{sec:cosmology}

In this section, we investigate homogeneous and isotropic cosmology in GDH theories based on the transformation property of the metric at both the background and perturbative levels.
In particular, we show that there is a nontrivial subclass of GDH theories where the propagation speed of gravitational waves coincides with that of light.
We also clarify the conditions for the absence of ghost/gradient instabilities for tensor and scalar perturbations.

\subsection{ADM decomposition under the unitary gauge}\label{ssec:ADM}

We first study how the Arnowitt-Deser-Misner (ADM) variables are transformed under the generalized disformal transformation.
This is useful in the context of cosmology where the scalar field has a timelike profile and hence defines a preferred time direction.
Note, however, that the results in this subsection apply to arbitrary spacetimes with a timelike scalar profile.

Let us introduce the ADM decomposition of the metric~$g_\mn$ as
    \be
    g_\mn \D x^\mu \D x^\nu=-\mN^2\D t^2+\ga_{ij}(\D x^i+\mN^i\D t)(\D x^j+\mN^j\D t),
    \ee
where $\mN$ is the lapse function, $\mN^i$ is the shift vector, and $\ga_{ij}$ is the induced metric.
Likewise, the ADM decomposition of the disformal metric~$\bar{g}_\mn$ in \eqref{disformal2} is written as
    \be
    \bar{g}_\mn \D x^\mu \D x^\nu=-\bar{\mN}^2\D t^2+\bar{\ga}_{ij}(\D x^i+\bar{\mN}^i\D t)(\D x^j+\bar{\mN}^j\D t).
    \ee
Under the unitary gauge where the scalar field is spatially uniform [i.e., $\phi=\phi(t)$], we have
    \be
    X=-\fr{\dot{\phi}^2}{\mN^2}, \qquad
    \bar{X}=-\fr{\dot{\phi}^2}{\bar{\mN}^2}, \label{X_ADM}
    \ee
with a dot denoting the derivative with respect to $t$.\footnote{In an epoch where the scalar field does not evolve monotonically in time, one cannot choose the unitary gauge.
Such a situation typically happens if we identify the scalar field as the inflaton in the context of standard inflationary cosmology, where the inflaton oscillates in the reheating era.
In this case, one should choose another appropriate gauge or perform a gauge-independent analysis.
Nevertheless, it is straightforward to redo our analysis either in another gauge or in a gauge-independent way, and there would be no conceptual problem.}
In what follows, we study the relationship between $(\mN,\mN_i,\ga_{ij})$ and $(\bar{\mN},\bar{\mN}_i,\bar{\ga}_{ij})$.

On performing the (generalized) disformal transformation, one could simultaneously redefine the time coordinate so that the new time-time component of the disformal metric coincides with the original one.
It is natural to perform such a coordinate redefinition when one works in a coordinate system where the lapse function is set to be unity, as done, e.g., in \cite{Minamitsuji:2014waa,Motohashi:2015pra}.
However, we do not do so in the present paper in order to clarify how each component of the metric is changed under the disformal transformation.
Hence, one should be careful about this issue when comparing the following results with those in earlier works.

For the transformation~\eqref{disformal2} to be invertible, we required that $\bar{X}$ should be written in the form~$\bar{X}(\phi,X)$, under which we obtain
    \be
    \bar{\mN}^2=-\fr{\dot{\phi}^2}{\bar{X}(\phi,-\dot{\phi}^2/\mN^2)}. \label{disformal_ADM1}
    \ee
Note that this equation can be solved for $\mN$ algebraically as
    \be
    \mN^2=-\fr{\dot{\phi}^2}{X(\phi,-\dot{\phi}^2/\bar{\mN}^2)}.
    \ee
Here, $X=X(\phi,\bar{X})$ is the inverse function of $\bar{X}=\bar{X}(\phi,X)$, which exists again due to the invertibility of the transformation~\eqref{disformal2} [see Eq.~\eqref{inv_cond}].
It is also straightforward to obtain the barred counterpart of the shift vector and the induced metric as
    \be
    \begin{split}
    \bar{\mN}_i&=\bar{g}_{0i}
    =F_0\mN_i+(F_2\dot{\phi}+F_3\dot{X})\Ds_iX, \\
    \bar{\gamma}_{ij}&=\bar{g}_{ij}
    =F_0\gamma_{ij}+F_3\Ds_iX\Ds_jX,
    \end{split} \label{disformal_ADM2}
    \ee
where $\Ds_i$ is the covariant derivative associated with $\ga_{ij}$.
The quantities~$Y$ and $Z$ are written as
    \be
    \begin{split}
    Y&=g^{0\mu}\dot{\phi}X_\mu
    =-\fr{\dot{\phi}}{\mN^2}\bra{\dot{X}-\mN^k\Ds_kX}, \\
    Z&=g^\mn X_\mu X_\nu
    =\Ds_kX\Ds^kX+\fr{Y^2}{X},
    \end{split} \label{YZ_ADM}
    \ee
so that
    \be
    W=Y^2-XZ=-X\Ds_kX\Ds^kX.
    \ee
Note that spatial indices for unbarred quantities are raised by $\gamma^{ij}$.
From this equation, we find that the quantity~$W$ does not contain $\dot{X}$ (and hence $\dot{\mN}$).
Using the above relations, we obtain the inverse spatial metric in the barred frame as
    \be
    \bar{\gamma}^{ij}=\fr{1}{F_0}\bra{\gamma^{ij}-\fr{XF_3}{XF_0-WF_3}\Ds^iX\Ds^jX},
    \ee
which is used to raise spatial indices for barred quantities.
For instance,
    \be
    \bar{\mN}^i=\bar{\gamma}^{ij}\bar{\mN}_j
    =\mN^i+\dot{\phi}\fr{XF_2+YF_3}{XF_0-WF_3}\Ds^iX.
    \ee

We stress that the invertibility of the generalized disformal transformation is crucial to obtain the transformation law for the lapse function in the simple form of \eqref{disformal_ADM1}.
Then, as it should be, the map~$(\mN,\mN_i,\ga_{ij})\mapsto (\bar{\mN},\bar{\mN}_i,\bar{\ga}_{ij})$ is invertible.
For noninvertible transformations that do not satisfy the condition~\eqref{Xbar_cond}, one can still obtain $\bar{\mN}$ as a functional of unbarred variables by employing \eqref{Xbar2}.
However, the transformation law now contains $\dot{\mN}$ and hence is no longer an algebraic equation for $\mN$.

\subsection{Background}\label{ssec:cosmo_BG}

Hereafter, we restrict ourselves to the spatially flat Friedmann-Lema{\^i}tre-Robertson-Walker (FLRW) background and perturbations about it.
The background metric and scalar field are described by
    \be
    g_\mn \D x^\mu \D x^\nu
    =-N(t)^2\D t^2+a(t)^2\delta_{ij}\D x^i \D x^j, \qquad
    \phi=\phi(t),
    \ee
with $a(t)$ being the scale factor. 
Then, the scalar-field kinetic term~$X$ also depends on $t$ only.
From \eqref{X_ADM} and \eqref{YZ_ADM}, the background value of the quantities~$X$, $Y$, and $Z$ are given by
    \be
    X_{\rm BG}\coloneqq -\fr{\dot{\phi}^2}{N^2}, \qquad
    Y_{\rm BG}\coloneqq \fr{X_{\rm BG}\dot{X}_{\rm BG}}{\dot{\phi}}, \qquad
    Z_{\rm BG}\coloneqq \fr{Y_{\rm BG}^2}{X_{\rm BG}},
    \label{XYZ_BG}
    \ee
respectively.
Here and in what follows, quantities with the subscript~``BG'' indicate their background value (and we omit the subscript when unnecessary).
Equation~\eqref{XYZ_BG} means that the quantity~$W=Y^2-XZ$ vanishes at the background level, which is an important property of the FLRW spacetime.

The barred-frame metric also takes the FLRW form,
    \begin{align}
    \bar{g}_\mn \D x^\mu \D x^\nu
    &=-\bar{N}(t)^2\D t^2+\bar{a}(t)^2\delta_{ij}\D x^i \D x^j,
    \end{align}
with $\bar{N}=\bar{N}(N)$ and $\bar{a}=\bar{a}[N,a]$ given by
    \be
    \fr{\bar{N}^2}{N^2}=\fr{X_{\rm BG}}{\bar{X}(\phi,X_{\rm BG})}=\fr{\mF_{\rm BG}}{F_{0,{\rm BG}}}, \qquad
    \bar{a}=F_{0,{\rm BG}}^{1/2}\,a.
    \label{barNa}
    \ee
Hence, on the FLRW background, $F_0>0$ and $\mF>0$ should be required in order to preserve the metric signature.
Note that we have used \eqref{Xbar2} with $W=0$ and also $\bar{X}=\bar{X}(\phi,X)$, which follows from the invertibility condition~\eqref{inv_cond}.
It should also be noted that \eqref{barNa} can be inverted to obtain $N=N(\bar{N})$ and $a=a[\bar{N},\bar{a}]$, which is consistent with the result of \cite{Babichev:2021bim}.

Let us now investigate perturbations about the FLRW background.
In principle, the formulae in \S\ref{ssec:ADM} provide transformation laws for nonlinear perturbations.
Actually, the authors of \cite{Motohashi:2015pra} studied how nonlinear perturbations about the FLRW background are transformed under conventional disformal transformations.
However, the extension of their discussion to generalized disformal transformations would be nontrivial due to the additional terms in the transformation laws, i.e., the second term of each equation in \eqref{disformal_ADM2}.
In any case, the formulae in \S\ref{ssec:ADM} are useful to study linear perturbations about the FLRW background, which we study in the following subsections.

\subsection{Tensor perturbations}\label{ssec:cosmo_tensor_pert}

We first consider tensor perturbations about the FLRW background.
In the unbarred frame, the tensor perturbations~$h_{ij}$ appear in the metric as
    \be
    g_\mn \D x^\mu \D x^\nu
    =-N^2\D t^2+a^2\bra{\delta_{ij}+h_{ij}}\D x^i \D x^j, \label{tensor_pert}
    \ee
with $h_{ii}=0=\pa_j h_{ij}$.
Note that $X$ remains unperturbed under the tensor perturbations.
Hence, performing the disformal transformation~\eqref{disformal2} on \eqref{tensor_pert}, we have
    \be
    \bar{g}_\mn \D x^\mu \D x^\nu
    =-\bar{N}^2\D t^2+\bar{a}^2\bra{\delta_{ij}+h_{ij}}\D x^i \D x^j.
    \ee
This means that the tensor perturbations are disformally invariant, i.e.,
    \be
    \bar{h}_{ij}=h_{ij}, \label{tensor_pert_disformal}
    \ee
which is consistent with the result in \cite{Minamitsuji:2021dkf}.

For Horndeski theories described by the action~\eqref{Horndeski}, the quadratic action for the tensor perturbations takes the form~\cite{Kobayashi:2011nu},
    \begin{align}
    S_T^{(2)}[h_{ij}]
    =\fr{1}{8}\int \D t\D^3x\,Na^3\brb{\fr{\mG_T}{N^2}\dot{h}_{ij}^2-\fr{\mF_T}{a^2}(\pa_k h_{ij})^2},
    \label{quad_action_tensor}
    \end{align}
where we have defined
    \be
    \begin{split}
    \mG_T&\coloneqq 2\brb{G_4-2XG_{4X}+\fr{X}{2}G_{5\phi}+H(-X)^{3/2}G_{5X}}, \\
    \mF_T&\coloneqq 2\bra{G_4-\fr{X}{2}G_{5\phi}+\fr{Y}{2}G_{5X}},
    \end{split} \label{GTFT}
    \ee
with $H\coloneqq \dot{a}/(Na)$ being the Hubble parameter.
So long as there is no tensor perturbation associated with the matter sector, one can use this quadratic action to study the linear dynamics of the tensor perturbations associated with the metric.
Here and in what follows, coefficients in the quadratic action are evaluated at the background.
Also, repeated spatial indices of perturbation variables are contracted by the Kronecker delta.
From the above action, we find the squared sound speed for the tensor perturbations as
    \be
    c_T^2=\fr{\mF_T}{\mG_T}.
    \label{cT_original}
    \ee
We see that the tensor perturbations are free of ghost/gradient instabilities if
    \be
    \mG_T>0, \qquad
    \mF_T>0. \label{stability_tensor_original}
    \ee

The quadratic action~\eqref{quad_action_tensor} for Horndeski theories can be straightforwardly mapped to the one for GDH theories.
More concretely, we first replace all the variables by barred ones and then translate it in the unbarred language by use of \eqref{barNa} and \eqref{tensor_pert_disformal}, i.e.,
    \begin{align}
    \ti{S}_T^{(2)}[h_{ij}]
    &=\fr{1}{8}\int \D t\D^3x\,\bar{N}\bar{a}^3\brb{\fr{\bar{\mG}_T}{\bar{N}^2}\dot{\bar{h}}_{ij}^2-\fr{\bar{\mF}_T}{\bar{a}^2}(\pa_k \bar{h}_{ij})^2} \nonumber \\
    &=\fr{1}{8}\int \D t\D^3x\,Na^3\mJ\brb{\fr{\ti{\mG}_T}{N^2}\dot{h}_{ij}^2-\fr{\ti{\mF}_T}{a^2}(\pa_k h_{ij})^2},
    \label{quad_action_tensor_GDH}
    \end{align}
where we have defined
    \be
    \begin{split}
    \bar{\mG}_T&\coloneqq 2\brb{G_4-2\bar{X}G_{4X}+\fr{\bar{X}}{2}G_{5\phi}+\bar{H}(-\bar{X})^{3/2}G_{5X}}, \\
    \bar{\mF}_T&\coloneqq 2\bra{G_4-\fr{\bar{X}}{2}G_{5\phi}+\fr{\bar{Y}}{2}G_{5X}},
    \end{split}
    \ee
with $\bar{H}\coloneqq \dot{\bar{a}}/(\bar{N}\bar{a})$ and
    \be
    \ti{\mG}_T\coloneqq \fr{\bar{X}}{X}\bar{\mG}_T, \qquad
    \ti{\mF}_T\coloneqq \fr{\bar{\mF}_T}{F_0}.
    \ee
Note that the functions~$G_4$ and $G_5$ as well as their derivatives are now evaluated at $(\phi,\bar{X})$.
The squared sound speed is given by
    \be
    \ti{c}_T^2=\fr{\ti{\mF}_T}{\ti{\mG}_T}
    =\fr{X\bar{\mF}_T}{\bar{X}F_0\bar{\mG}_T}.
    \label{cT_GDH}
    \ee
Also, the conditions for the absence of ghost/gradient instabilities are given by
    \be
    \ti{\mG}_T>0, \qquad
    \ti{\mF}_T>0. \label{stability_tensor_GDH}
    \ee

In what follows, we discuss issues on the speed of gravitational waves.
First, it should be noted that we always assume the existence of matter field(s) which are minimally coupled, and hence light propagates at a unit speed in both the seed Horndeski and GDH theories.
As mentioned earlier, the almost simultaneous detection of the gravitational-wave event~GW170817 and the gamma-ray burst~170817A posed a tight constraint on the propagation speed of gravitational waves in the late-time universe, which motivates us to study theories where gravitational waves propagate at a unit speed.
It is now well known that Horndeski theories with $G_{4X}=G_5=0$ satisfy $c_T^2=1$ regardless of the background evolution so long as we consider a homogeneous and isotropic background~\cite{Creminelli:2017sry,Ezquiaga:2017ekz,Baker:2017hug}.
Indeed, we have $\mG_T=\mF_T=2G_4$ in this case, and hence $c_T^2=1$ is deduced from \eqref{cT_original}.
A subclass of DHOST theories with $c_T^2=1$ is also isolated~\cite{Langlois:2017dyl}.

Here, we further enlarge the viable theory space by using the framework of GDH theories.
Let us use Horndeski theories with $G_5=0$ (and not necessarily $G_{4X}=0$) as a seed, for which
    \be
    \bar{\mG}_T=2\brb{G_4(\phi,\bar{X})-2\bar{X}G_{4X}(\phi,\bar{X})}, \qquad
    \bar{\mF}_T=2G_4(\phi,\bar{X}).
    \ee
Note that both $\bar{\mG}_T$ and $\bar{\mF}_T$ can be translated into functions of $(\phi,X)$ since we are interested in the invertible subclass of the generalized disformal transformations for which $\bar{X}=\bar{X}(\phi,X)$.
The reason why we set $G_5=0$ is to remove the term containing the scale factor~$a$, whose time evolution depends on the matter sector.
Then, the squared sound speed of tensor perturbations in \eqref{cT_GDH} can be written as
    \be
    \ti{c}_T^2=\fr{XG_4(\phi,\bar{X})}{\bar{X}F_0\brb{G_4(\phi,\bar{X})-2\bar{X}G_{4X}(\phi,\bar{X})}}.
    \ee
Therefore, if we choose the conformal factor of the disformal transformation as\footnote{Since $W_{\rm BG}=Y_{\rm BG}^2-X_{\rm BG}Z_{\rm BG}=0$ on the FLRW background [see Eq.~\eqref{XYZ_BG}], this requirement does not completely fix the functional form of $F_0(\phi,X,Y,Z)$. This ambiguity may be solved by studying a spatially inhomogeneous background (e.g., black hole solutions) where $W_{\rm BG}\ne 0$.}
    \be
    F_0=\fr{XG_4(\phi,\bar{X})}{\bar{X}\brb{G_4(\phi,\bar{X})-2\bar{X}G_{4X}(\phi,\bar{X})}}, \label{F0_GDH}
    \ee
we can make $\ti{c}_T^2=1$.
This also means that one can map a Horndeski model with $c_T^2\ne 1$ to one with $\ti{c}_T^2=1$ by generalized disformal transformation, which is a generalization of the results in \cite{Creminelli:2017sry,Ezquiaga:2017ekz,Babichev:2017lmw}.
We note that the above expression for $F_0$ does not contain either $Y$ or $Z$ (though there could be an implicit dependence through $W$).
As a result, $\mJ=F_0\mF^{1/2}$ is also independent of $Y$ or $Z$ on the FLRW background, which can be verified by use of \eqref{barNa}.
As we shall see in \S\ref{ssec:coupling_linear}, this property leads to the consistency of matter coupling at the level of linear cosmological perturbations.

One can also see that the expression of the speed of gravitational waves remains unchanged for transformations with
    \be
    F_0=\fr{X}{\bar{X}}, \label{cT_unchanged}
    \ee
or equivalently, $F_1 X+2F_2Y+F_3Z=0$.
This is the case for purely conformal transformations.
For conventional disformal transformations, this requirement yields $F_1=0$, i.e., the transformation must be conformal.
On the other hand, non-conformal transformations satisfying the condition~\eqref{cT_unchanged} exist within generalized disformal transformations.
However, one should bear in mind that, although the expression of the gravitational wave speed itself does not change under the disformal transformation if the condition~\eqref{cT_unchanged} is satisfied, its value could change.
This is because $c_T^2$ and $\ti{c}_T^2$ depend on the time evolution of $\phi$ in general, which could change under the disformal transformation.

\subsection{Vector perturbations}\label{ssec:cosmo_vector_pert}

In this subsection, we study vector perturbations about the FLRW background.
In the unbarred frame, the vector perturbations appear in the metric as
    \be
    g_\mn \D x^\mu \D x^\nu
    =-N^2\D t^2+2NaU_i\D t\D x^i+a^2\bra{\delta_{ij}+2\pa_{(i}V_{j)}}\D x^i \D x^j,
    \ee
with $\pa_iU_i=0=\pa_iV_i$.
Similarly, we write the barred-frame metric with vector perturbations as
    \be
    \bar{g}_\mn \D x^\mu \D x^\nu
    =-\bar{N}^2\D t^2+2\bar{N}\bar{a}\bar{U}_i\D t\D x^i+\bar{a}^2\bra{\delta_{ij}+2\pa_{(i}\bar{V}_{j)}}\D x^i \D x^j,
    \ee
with $\pa_i\bar{U}_i=0=\pa_i\bar{V}_i$.
The ADM variables are given by 
    \be
    \begin{split}
    &\mN=N\bra{1+\fr{1}{2}U_i^2+\cdots}, \qquad 
    \mN_i=NaU_i, \qquad \gamma_{ij}=a^2\bra{\delta_{ij}+2\pa_{(i}V_{j)}}, \\
    &\bar{\mN}=\bar{N}\bra{1+\fr{1}{2}\bar{U}_i^2+\cdots}, \qquad 
    \bar{\mN}_i=\bar{N}\bar{a}\bar{U}_i, \qquad \bar{\gamma}_{ij}=\bar{a}^2\bra{\delta_{ij}+2\pa_{(i}\bar{V}_{j)}},
    \end{split} \label{ADM_vector-pert}
    \ee
where the ellipses denote terms of third or higher order in perturbations.
We remind that $U_i^2$ means $\delta_{ij}U_i U_j$.
Note that $X$ remains unperturbed under the vector perturbations up to the linear order.
Then, \eqref{disformal_ADM1} and \eqref{disformal_ADM2} allow us to express barred variables in terms of unbarred ones as
    \be
    \bar{\mN}^2\simeq \bar{N}^2, \qquad
    \bar{\mN}_i\simeq \bra{\fr{\bar{X}F_0}{X}}^{1/2}\bar{N}\bar{a}U_i, \qquad
    \bar{\gamma}_{ij}\simeq \bar{a}^2\bra{\delta_{ij}+2\pa_{(i}V_{j)}},
    \ee
where the symbol~$\simeq$ denotes the equality up to the linear order in perturbations.
Comparing these with the second line of \eqref{ADM_vector-pert}, we find the following relation between the linear vector perturbations in the barred and unbarred frames:
    \be
    \bar{U}_i=\bra{\fr{\bar{X}F_0}{X}}^{1/2}U_i, \qquad
    \bar{V}_i=V_i.
    \ee

Note that the vector modes are nondynamical in Horndeski theories and hence in GDH theories, unless the matter sector introduces a dynamical vector mode.

\subsection{Scalar perturbations}\label{ssec:cosmo_scalar_pert}

Let us now study scalar perturbations about the FLRW background.
We focus on the gravitational sector in this section and the coupling with the matter sector is studied in \S\ref{sec:coupling}.
We take the unitary gauge where $\phi=\phi(t)$ and consider only the metric perturbations.
In the unbarred frame, the scalar perturbations appear in the metric as
    \be
    g_\mn \D x^\mu \D x^\nu
    =-N^2(1+2\alpha)\D t^2+2Na\pa_i\chi\D t\D x^i+a^2\brb{(1+2\zeta)\delta_{ij}+2\pa_i\pa_jE}\D x^i \D x^j,
    \ee
where $\alpha$, $\chi$, $\zeta$, and $E$ denote the perturbation variables.
Likewise, in the barred frame,
    \be
    \bar{g}_\mn \D x^\mu \D x^\nu
    \coloneqq -\bar{N}^2(1+2\bar{\alpha})\D t^2+2\bar{N}\bar{a}\pa_i\bar{\chi}\D t\D x^i+\bar{a}^2\brb{(1+2\bar{\zeta})\delta_{ij}+2\pa_i\pa_j\bar{E}}\D x^i \D x^j.
    \ee
In terms of the ADM variables,
    \be
    \begin{split}
    &\mN=N\brb{1+\alpha-\fr{\alpha^2}{2}+\fr{1}{2}(\pa_i\chi)^2+\cdots}, \qquad 
    \mN_i=Na\pa_i\chi, \qquad \gamma_{ij}=a^2\brb{(1+2\zeta)\delta_{ij}+2\pa_i\pa_jE}, \\
    &\bar{\mN}=\bar{N}\brb{1+\bar{\alpha}-\fr{\bar{\alpha}^2}{2}+\fr{1}{2}(\pa_i\bar{\chi})^2+\cdots}, \qquad 
    \bar{\mN}_i=\bar{N}\bar{a}\pa_i\bar{\chi}, \qquad \bar{\gamma}_{ij}=\bar{a}^2\brb{(1+2\bar{\zeta})\delta_{ij}+2\pa_i\pa_j\bar{E}},
    \end{split}
    \ee
where the ellipses denote terms of third or higher order in perturbations.
Up to the linear order in perturbations, we have
    \be
    X\simeq X_{\rm BG}(1-2\alpha), \qquad
    Y\simeq Y_{\rm BG}(1-4\alpha)-\fr{2X^2}{\dot{\phi}}\dot{\alpha}, \qquad
    Z\simeq Z_{\rm BG}(1-6\alpha)-\fr{4XY}{\dot{\phi}}\dot{\alpha}.
    \ee
Hence, functions of $(\phi,X,Y,Z)$ are also perturbed, e.g.,
    \begin{align}
    F_0(\phi,X,Y,Z)
    \simeq F_{0,{\rm BG}}-2(XF_{0X}+2YF_{0Y}+3ZF_{0Z})\alpha-\fr{2X}{\dot{\phi}}(XF_{0Y}+2YF_{0Z})\dot{\alpha}. \label{expansion_F0}
    \end{align}
With these relations, \eqref{disformal_ADM1} and \eqref{disformal_ADM2} yield
    \be
    \begin{split}
    \bar{\mN}^2&\simeq \bar{N}^2\bra{1+\fr{2X\bar{X}_{X}}{\bar{X}}\alpha}, \\
    \bar{\mN}_i&\simeq \bra{\fr{\bar{X}F_0}{X}}^{1/2}\bar{N}\bar{a} \pa_i \chi-2\dot{\phi}(XF_2+YF_3)\pa_i\alpha, \\
    \bar{\gamma}_{ij}&\simeq F_0a^2\brb{(1+2\zeta)\delta_{ij}+2\pa_i\pa_jE}.
    \end{split}
    \ee
Note that, in the expression for $\bar{\gamma}_{ij}$, one should keep the linear term of $F_0$ shown in \eqref{expansion_F0} since it is multiplied by the background part.
Therefore, the linear scalar perturbations in the barred frame are related to those in the unbarred frame as
    \be
    \begin{split}
    \bar{\alpha}&=\fr{X\bar{X}_{X}}{\bar{X}}\alpha, \\
    \bar{\chi}&=\bra{\fr{\bar{X}F_0}{X}}^{1/2}\brb{\chi-\fr{2\dot{\phi}}{F_0Na}(XF_2+YF_3)\alpha}, \\
    \bar{\zeta}&=\zeta-\fr{1}{F_0}(XF_{0X}+2YF_{0Y}+3ZF_{0Z})\alpha-\fr{X}{\dot{\phi}F_0}(XF_{0Y}+2YF_{0Z})\dot{\alpha}, \\
    \bar{E}&=E.
    \end{split} \label{scalar_pert_disformal}
    \ee
Thanks to the invertibility condition~\eqref{inv_cond} imposed on the nonlinear transformation law~\eqref{disformal2}, the invertibility is manifest also at the level of linear perturbations.
Note that, without the condition~\eqref{Xbar_cond}, the time derivative of $\alpha$ shows up in $\bar{\alpha}$, which makes the transformation of the lapse perturbation noninvertible.
It should also be noted that the above results are consistent with those in \cite{Minamitsuji:2021dkf}.

One can discuss the linear stability of scalar perturbations in a similar manner as in the case of tensor perturbations studied in \S\ref{ssec:cosmo_tensor_pert}.
Namely, we first start from the quadratic action for the scalar perturbations in Horndeski theories, and then map it to the one for GDH theories.
As mentioned earlier, we ignore matter field(s) in this section and we shall discuss the case with matter field(s) in \S\ref{sec:coupling}.
In what follows, we choose the gauge where $E=0$ (on top of $\delta\phi=0$).
Note that this is a complete gauge fixing and hence can be imposed at the Lagrangian level~\cite{Motohashi:2016prk}.
For Horndeski theories described by the action~\eqref{Horndeski}, the perturbation variables~$\alpha$ and $\chi$ (i.e., those associated with the lapse function and the shift vector, respectively) are auxiliary variables, and hence $\zeta$ plays the role of master variable.
The quadratic action can be written as~\cite{Kobayashi:2011nu}
    \begin{align}
    S_S^{(2)}[\zeta]
    =\int \D t\D^3x\,Na^3\brb{\fr{\mG_S}{N^2}\dot{\zeta}^2-\fr{\mF_S}{a^2}(\pa_k \zeta)^2},
    \label{quad_action_scalar}
    \end{align}
where we have defined
    \be
    \mG_S\coloneqq \fr{\mG_T}{\Theta^2}\bra{3\Theta^2+\mG_T\Sigma}, \qquad
    \mF_S\coloneqq \fr{1}{Na}\bra{\fr{a}{\Theta}\mG_T^2}^{\boldsymbol{\cdot}}-\mF_T,
    \label{GSFS}
    \ee
with
    \be
    \begin{split}
    \Sigma&\coloneqq XG_{2X}+2X^2G_{2XX}
    -6H\dot{\phi}\bra{2XG_{3X}+X^2G_{3XX}}-X\bra{G_{3\phi}+XG_{3\phi X}} \\
    &\quad -6H^2\bra{G_4-7XG_{4X}-16X^2G_{4XX}-4X^3G_{4XXX}}-6H\dot{\phi}\bra{G_{4\phi}+5XG_{4\phi X}+2X^2G_{4\phi XX}} \\
    &\quad +2H^3\dot{\phi}\bra{15XG_{5X}+13X^2G_{5XX}+2X^3G_{5XXX}}+3H^2X\bra{6G_{5\phi}+9XG_{5\phi X}+2X^2G_{5\phi XX}}, \\
    \Theta&\coloneqq \dot{\phi}XG_{3X}+2H\bra{G_{4}-4XG_{4X}-4X^2G_{4XX}}+\dot{\phi}\bra{G_{4\phi}+2XG_{4\phi X}}-H^2\dot{\phi}\bra{5XG_{5X}+2X^2G_{5XX}} \\
    &\quad -HX\bra{3G_{5\phi}+2XG_{5\phi X}}.
    \end{split} \label{SigmaTheta}
    \ee
We now map the action~\eqref{quad_action_scalar} by replacing all the variables by barred ones and then translate it in the unbarred language by use of \eqref{barNa}, i.e.,
    \begin{align}
    \ti{S}_S^{(2)}[\bar{\zeta}]
    &=\int \D t\D^3x\,\bar{N}\bar{a}^3\brb{\fr{\bar{\mG}_S}{\bar{N}^2}\dot{\bar{\zeta}}^2-\fr{\bar{\mF}_S}{\bar{a}^2}(\pa_k \bar{\zeta})^2} \nonumber \\
    &=\int \D t\D^3x\,Na^3\mJ\brb{\fr{\ti{\mG}_S}{N^2}\dot{\bar{\zeta}}^2-\fr{\ti{\mF}_S}{a^2}(\pa_k \bar{\zeta})^2},
    \label{quad_action_scalar_GDH}
    \end{align}
where $\bar{\mG}_S$ and $\bar{\mF}_S$ are the barred counterparts of $\mG_S$ and $\mF_S$, and
    \be
    \ti{\mG}_S\coloneqq \fr{\bar{X}}{X}\bar{\mG}_S, \qquad
    \ti{\mF}_S\coloneqq \fr{\bar{\mF}_S}{F_0}.
    \ee
Therefore, the conditions for the absence of ghost/gradient instabilities are given by
    \be
    \ti{\mG}_S>0, \qquad
    \ti{\mF}_S>0. \label{stability_scalar_GDH}
    \ee
Note that the master variable~$\zeta$ is transformed in a nontrivial manner under the generalized disformal transformation.
As a result, the master variable in GDH theories, i.e., $\bar{\zeta}$, is a linear combination of $\zeta$, $\alpha$, and $\dot{\alpha}$ [see Eq.~\eqref{scalar_pert_disformal}].
Related to this point, when reconstructing metric perturbations from the master variable, one also needs the transformation law~\eqref{scalar_pert_disformal} for each perturbation variable.

\section{Coupling with a matter field}\label{sec:coupling}

By construction, any DHOST theories have three physical DOFs in total:
two from the metric and one from the scalar field.
Let us consider matter field(s) coupled to DHOST theories.
Provided that the matter sector has $n$ DOFs, one would expect that the whole system has $n+3$ DOFs.
However, this is not the case in general:
The whole system can have more than $n+3$ DOFs, which happens when the matter sector does not respect the degeneracy conditions imposed on the gravitational sector (see \cite{Deffayet:2020ypa} for a detailed discussion).
One should avoid such a situation as the revived DOF(s) would be unstable Ostrogradsky mode(s).
When a DHOST theory with matter field(s) has exactly $n+3$ DOFs, the matter coupling is said to be consistent.

Regarding this point, the authors of \cite{Deffayet:2020ypa} investigated which matter fields can be coupled to DH theories in a consistent manner.
Along this line of thought, in this section, we shall clarify which GDH theories 
can accommodate consistent matter coupling.
We perform calculations in the ``Horndeski frame'' where the matter scalar field is coupled to the generalized disformal metric.
As we shall see below, this not only simplifies the analysis but also clarifies the origin of the revival of the Ostrogradsky mode in the presence of a matter field.
Note that, throughout this section, functions of $(\phi,X,Y,Z)$ are regarded as those of $(\phi,X,Y,W)$, i.e., $W=Y^2-XZ$ is regarded as an independent variable instead of $Z$ for the reason explained below.

\subsection{Linear cosmological perturbations}
\label{ssec:coupling_linear}

In order to see whether or not the generalized disformal coupling in the matter sector leads to an unwanted extra DOF, let us first consider cosmological scalar perturbations.
As in \S\ref{ssec:cosmo_scalar_pert}, we impose the gauge conditions~$E=\delta\phi=0$ which completely fix the gauge DOFs.
For concreteness, in this section, we consider only a k-essence scalar field~\cite{ArmendarizPicon:1999rj} as a choice of the matter sector, which is often used in the context of cosmology to mimic a barotropic perfect fluid.
Nevertheless, we shall see that our conclusion actually does not rely on the details of the matter Lagrangian and applies to a much broader class of matter Lagrangian.
Let $\sigma$ denote the k-essence scalar field (whose perturbation is denoted by $\delta\sigma$) and consider the following Horndeski-frame action:
    \be
    S=S_{\rm H}[g_\mn,\phi]+S_{\rm m}[\bar{g}_\mn,\sigma],
    \ee
where $S_{\rm H}$ denotes the action of Horndeski theories~\eqref{Horndeski} and the matter action~$S_{\rm m}$ is given by
    \be
    S_{\rm m}[\bar{g}_\mn,\sigma]\coloneqq 
    \int \D^4x\sqrt{-\bar{g}}\,P(\kinmat), \qquad
    \kinmat\coloneqq \bar{g}^{\alpha\beta}\pa_\alpha\sigma\pa_\beta\sigma.
    \label{matter_action}
    \ee
Here, $\bar{g}_\mn$ is the generalized disformal metric defined in \eqref{disformal2}.

The quadratic Lagrangian has contributions from both the gravitational and matter sectors.
The contribution from the gravitational sector~$\mL^{(2)}_{\rm H}$ is given by~\cite{Kobayashi:2011nu}\footnote{Terms with $(\pa^2\chi)^2$ vanish after using the Hamiltonian constraint.
Also, if one ignores the matter sector, integrating out $\alpha$ and $\chi$ from \eqref{qLag_H} yields the quadratic action~\eqref{quad_action_scalar}.}
    \be
    \mL^{(2)}_{\rm H}[\alpha,\chi,\zeta]=
    Na^3\bra{-3\mG_T\fr{\dot{\zeta}^2}{N^2}-\fr{\mF_T}{a^2}\zeta\pa^{2}\zeta+\Sigma\alpha^2-\fr{2\Theta}{a^2}\alpha\pa^{2}\chi+6\Theta\alpha\fr{\dot{\zeta}}{N}-\fr{2\mG_T}{a^2}\alpha\pa^{2}\zeta+\fr{2\mG_T}{a^2}\fr{\dot{\zeta}}{N}\pa^{2}\chi}, \label{qLag_H}
    \ee
with $\mG_T$, $\mF_T$, $\Sigma$, and $\Theta$ being those defined in \eqref{GTFT} and \eqref{SigmaTheta}.
Note that $\mL^{(2)}_{\rm H}$ does not contain $\dot{\alpha}$, i.e., the time derivative of the lapse perturbation.
On the other hand, the contribution from the matter sector~$\mL^{(2)}_{\rm m}$ contains $\dot{\alpha}$, which makes $\alpha$ dynamical and hence leads to an inconsistent matter coupling in general.
In order to avoid this problem, we require that the kinetic matrix of the matter sector is degenerate.
The kinetic terms in $\mL^{(2)}_{\rm m}$ are given by
    \be
    \mL^{(2)}_{\rm m}\supset
    Na^3\mJ\bra{\fr{2X^4}{\dot{\phi}^2}\fr{\mJ_{YY}}{\mJ}P\dot{\alpha}^2+\fr{4X^2\dot{\sigma}}{\dot{\phi}\bar{N}^2}\fr{\mJ_Y}{\mJ} P'\dot{\alpha}\dot{\delta\sigma}-\fr{P'+2\kinmat P''}{\bar{N}^2}\dot{\delta\sigma}^2}, \label{qlag_matter}
    \ee
where $P'\coloneqq \D P/\D \kinmat$ and $P''\coloneqq \D^2 P/\D \kinmat^2$, and we recall that $\mJ=F_0\mF^{1/2}$.
Note that, as mentioned earlier, we now regard $\mJ$ as a function of $(\phi,X,Y,W)$ so that $\dot{\alpha}$ appears only through $Y$.
Since we require that the kinetic matrix associated with \eqref{qlag_matter} should be degenerate for any function~$P$, the quantity~$\mJ$ should satisfy
    \be
    \mJ_Y=\mJ_{YY}=0, \label{consistent_matter_coupling}
    \ee
at the background level.
Interestingly, GDH theories with $\ti{c}_T^2=1$ obtained in \S\ref{ssec:cosmo_tensor_pert} automatically satisfy this condition [see Eq.~\eqref{F0_GDH} and the subsequent discussion].

\subsection{Nonlinear and background-independent consideration}
\label{ssec:coupling_nonlinear}

The above discussion at the linear level suggests that the appearance of the time derivative of the lapse function results in an extra DOF.
With the formulae in \S\ref{ssec:ADM}, one can obtain the condition on the generalized disformal transformation under which the matter action~\eqref{matter_action} does not contain $\dot{\mN}$ even at the nonlinear level on an arbitrary background with a timelike scalar profile.
We note that $\dot{\mN}$ can show up only through the components of the disformal metric~$\bar{\mN}_i$ and $\bar{\gamma}_{ij}$ because the barred lapse function~$\bar{\mN}$ does not depend on $\dot{\mN}$ under the invertibility condition~\eqref{Xbar_cond} [see Eq.~\eqref{disformal_ADM1}].
From \eqref{disformal_ADM2} and \eqref{YZ_ADM}, we have
    \be
    \bar{\mN}_i
    =F_0\mN_i+\brb{\fr{\dot{\phi}}{X}(XF_2+YF_3)+F_3\mN^k\Ds_kX}\Ds_iX, \qquad
    \bar{\gamma}_{ij}
    =F_0\gamma_{ij}+F_3\Ds_iX\Ds_jX.
    \ee
As in the previous subsection, it is useful to regard all functions of $(\phi,X,Y,Z)$ as functions of $(\phi,X,Y,W)$ so that $\dot{\mN}$ appears only through $Y$.
Then, requiring the $Y$-independence of $\bar{\mN}_i$ and $\bar{\gamma}_{ij}$, we obtain the following condition:
    \be
    F_{0Y}=F_{3Y}=XF_{2Y}+F_3=0.
    \label{no_dot_N}
    \ee
Of course, on top of this condition, one has to require the invertibility condition~\eqref{inv_cond}.
Since we are now regarding $W$ as an independent variable instead of $Z$, the invertibility condition~\eqref{inv_cond} reads
    \be
    F_0\ne 0, \qquad
    \mF\ne 0, \qquad
    \bar{X}_Y=\bar{X}_W=0, \qquad
    \bar{X}_X\ne 0, \qquad
    \left|\fr{\pa(\bar{Y},\bar{W})}{\pa(Y,W)}\right|\ne 0.
    \label{inv_cond_W}
    \ee
It should also be noted that, under the invertibility condition, the condition~\eqref{no_dot_N} is strictly stronger than the condition~\eqref{consistent_matter_coupling}.
Actually, GDH theories with $\ti{c}_T^2=1$ do not necessarily satisfy \eqref{no_dot_N}.
Thus, imposing the conditions~\eqref{no_dot_N} and \eqref{inv_cond_W} isolates a subclass of GDH theories with $\ti{c}_T^2=1$ where the matter coupling would be consistent even at the nonlinear level on an arbitrary background spacetime with a timelike scalar profile.

One can write down an explicit form of invertible generalized transformations that satisfy the conditions~\eqref{no_dot_N} and \eqref{inv_cond_W} for consistent matter coupling.\footnote{The authors would like to thank Atsushi Naruko, Ryo Saito, Norihiro Tanahashi, and Daisuke Yamauchi for a correspondence on this point.
The following arguments are based on the correspondence with them and show that our result is consistent with theirs in \cite{Naruko:2022vuh} under the invertibility condition~\eqref{inv_cond_W}.}
Equation~\eqref{no_dot_N} fixes the $Y$-dependence of $F_0$, $F_2$, $F_3$ as
    \be
    F_0=F_0(\phi,X,W)\,(\ne 0), \qquad
    F_3=F_3(\phi,X,W), \qquad
    F_2=h(\phi,X,W)-\fr{Y}{X}F_3(\phi,X,W),
    \label{GDT_consistent_F023}
    \ee
with $h$ being an arbitrary function of $(\phi,X,W)$.
Then, from \eqref{Xbar2} and the third condition in \eqref{inv_cond_W}, the functional form of $F_1$ can be fixed as
    \be
    F_1=\fr{\mF-Wh^2}{\bar{X}\mF}-\fr{F_0}{X}-\fr{2Y}{X}h+\fr{Y^2}{X^2}F_3, \qquad
    \mF=\fr{XF_0-WF_3}{\bar{X}}\,(\ne 0), \label{GDT_consistent_F1}
    \ee
where we regard $\bar{X}=\bar{X}(\phi,X)$ as given and require $\bar{X}_X\ne 0$.
Also, by use of \eqref{YZbar} and \eqref{Wbar}, we have
    \be
    \bar{Y}=\bar{X}_X\bra{\fr{\bar{X}}{X}Y+\fr{Wh}{\mF}}+\bar{X}_\phi\bar{X}, \qquad
    \bar{W}=\fr{\bar{X}_X^2}{\mF}W,
    \ee
and hence the last condition in \eqref{inv_cond_W} reads $\bar{W}_W\ne 0$, or equivalently,
    \be
    \fr{\pa}{\pa W}\bra{\fr{XF_0-WF_3}{W}}\ne 0.
    \ee
To summarize, for given functions~$F_0(\phi,X,W)$, $F_3(\phi,X,W)$, $h(\phi,X,W)$, and $\bar{X}(\phi,X)$ such that
    \be
    F_0\ne 0, \qquad
    XF_0-WF_3\ne 0, \qquad
    \bar{X}_X\ne 0, \qquad
    \fr{\pa}{\pa W}\bra{\fr{XF_0-WF_3}{W}}\ne 0, \label{inv_cond_consistent}
    \ee
the generalized disformal transformation with the coefficient functions given by \eqref{GDT_consistent_F023} and \eqref{GDT_consistent_F1} is invertible and presumably accommodates consistent matter coupling under the unitary gauge.

Finally, we stress that the discussion in this section does not rely on the details of the matter Lagrangian:
So long as the matter Lagrangian~$\mL_{\rm m}[\bar{g}_\mn,\Psi]$ does not contain derivatives of $\bar{g}_\mn$, the time derivative of the lapse function does not arise in the matter sector, and hence the matter coupling would be consistent.
Therefore, the matter action does not have to be the k-essence one~\eqref{matter_action}.
For instance, one could incorporate a potential term into the k-essence action.
Also, the matter field can be standard gauge (or Proca) fields.
However, the situation is nontrivial for fermionic matter fields:
In the case of fermions, the matter Lagrangian is written in terms of not the metric itself but the tetrad, and hence one needs to develop the transformation law for the tetrad~\cite{Takahashi:2022ctx}, which is beyond the scope of the present paper.

\section{Conclusions}\label{sec:conc}

Disformal transformations have played an important role in the context of scalar-tensor theories.
In particular, invertible disformal transformations can be used to enlarge the framework of scalar-tensor theories free from Ostrogradsky ghosts.
Indeed, using the Horndeski class as a seed and performing conventional disformal transformations containing up to the first derivative of the scalar field on it, one can generate a more general class of ghost-free scalar-tensor theories, which belongs to the quadratic/cubic DHOST class.
A good thing is that the resultant class of theories, which we dubbed the disformal Horndeski (DH) class, can accommodate stable and viable cosmological solutions, whereas the remaining part of the quadratic/cubic DHOST class does not allow viable cosmology.  
Motivated by this fact, in the present paper, we performed a higher-derivative extension of invertible disformal transformations developed in \cite{Takahashi:2021ttd} on Horndeski theories to construct a novel class of ghost-free scalar-tensor theories, which we dubbed generalized disformal Horndeski (GDH) theories. 
While conventional ghost-free scalar-tensor theories contain derivatives of the scalar field up to the second order, our GDH theories contain the third derivative of the scalar field (see \S\ref{ssec:disformal_Horndeski}).

In \S\ref{sec:cosmology}, we studied linear cosmological perturbations in GDH theories.
We discussed how the perturbation variables are transformed under generalized disformal transformations and showed that the invertibility is manifest at the level of linear perturbations.
We also clarified the conditions for the absence of ghost/gradient instabilities for tensor and scalar perturbations.
Moreover, we identified a subclass of theories with $c_T^2=1$ (i.e., gravitational waves propagate at the speed of light), which is consistent with the almost simultaneous detection of the gravitational-wave event~GW170817 and the gamma-ray burst~170817A.

In \S\ref{sec:coupling}, we studied conditions under which a matter field can be consistently coupled to GDH theories without introducing unwanted extra DOF(s).
For simplicity, we worked in the frame where the gravitational action is given by the Horndeski one (and hence the matter field is coupled to the generalized disformal metric).
We found that the generalized disformal coupling in the matter sector introduces the time derivative of the lapse function in general, which results in an unwanted extra DOF.
Nevertheless, there is a subclass of GDH theories for which the matter coupling would be consistent at the level of linear cosmological perturbations (see \S\ref{ssec:coupling_linear}).
Interestingly, theories with $c_T^2=1$ belongs to this subclass.
Moreover, in \S\ref{ssec:coupling_nonlinear}, we specified a class of invertible generalized disformal transformations in which the time derivative of the lapse function does not show up.
The class of GDH theories obtained via such transformations would accommodate consistent matter coupling even at the nonlinear level on an arbitrary background spacetime, provided that the matter Lagrangian does not contain the derivative of the metric to which the matter field is minimally coupled.

There are several possible future directions.
It would be intriguing to study how GDH theories are embedded in 
the framework of effective field theory of inflation/dark energy.
This would also be helpful to specify the class of GDH theories without gravitational wave decay into the scalar field.
It is also interesting to investigate the screening mechanism in GDH theories.
As is well known, the Vainshtein screening mechanism~\cite{Vainshtein:1972sx,Babichev:2013usa} works successfully in Horndeski theories~\cite{DeFelice:2011th,Koyama:2013paa,Kase:2013uja}.
On the other hand, in conventional DHOST theories (i.e., DH theories), that lie outside the Horndeski class, it is known that the Vainshtein screening is broken inside astrophysical bodies~\cite{Kobayashi:2014ida,Koyama:2015oma,Saito:2015fza,Crisostomi:2017lbg,Langlois:2017dyl,Dima:2017pwp}.
The situation can be more nontrivial in GDH theories due to novel higher-derivative interactions, for which a detailed analysis is required.
Another interesting thing is to study the consistency of matter coupling without imposing the unitary gauge.
Actually, we precluded the revival of the Ostrogradsky ghost only under the unitary gauge, and hence there remains a shadowy mode which satisfies an elliptic differential equation on a spacelike hypersurface.
A shadowy mode itself is harmless but needs a careful treatment~\cite{DeFelice:2018mkq,DeFelice:2021hps}, so it would be useful to specify a subclass of GDH theories where matter fields can be coupled without introducing a shadowy mode.
Yet another is to investigate the singular (or noninvertible) subclass of generalized disformal transformations.
Within conventional disformal transformations, singular transformations have been studied in the context of mimetic gravity~\cite{Chamseddine:2013kea}.
Interestingly, if one starts from some seed scalar-tensor theory and performs a singular disformal transformation on it, then the resultant theory obtains an extra symmetry associated with the singular nature of the transformation~\cite{Takahashi:2017pje,Langlois:2018jdg}.
Namely, one can systematically construct DHOST theories from generically nondegenerate theories by use of singular transformations.
As we discuss in Appendix~\ref{AppC}, singular generalized disformal transformations with $\bar{X}_X=0$ can always be recast into the singular conformal transformation by use of an appropriate invertible transformation.
However, this does not exhaust all singular transformations and we need a further analysis.
We leave these issues for future studies.


\vskip5mm
{\bf Note added}: While we were finalizing this paper, we were informed of a research~\cite{Naruko:2022vuh} by Atsushi Naruko, Ryo Saito, Norihiro Tanahashi, and Daisuke Yamauchi on a similar subject.
Their results are consistent with ours on overlapping parts.
We would like to thank them for their kind correspondence.

\acknowledgments{
K.T.~would like to thank Takashi Hiramatsu, Shin'ichi Hirano, Tsutomu Kobayashi, Teruaki Suyama, Hiroaki W.~H.~Tahara, and Masahide Yamaguchi for useful discussions and constructive comments.
K.T.~was supported by JSPS (Japan Society for the Promotion of Science) KAKENHI Grant No.~JP21J00695.
M.M.~was supported by the Portuguese national fund through the Funda\c{c}\~{a}o para a Ci\^encia e a Tecnologia in the scope of the framework of the Decree-Law 57/2016 of August 29, changed by Law 57/2017 of July 19, and the Centro de Astrof\'{\i}sica e Gravita\c c\~ao through the Project~No.~UIDB/00099/2020.
M.M.~also would like to thank Yukawa Institute for Theoretical Physics for their hospitality under the Visitors Program of FY2022.
H.M.~was supported by JSPS KAKENHI Grants No.~JP18K13565 and No.~JP22K03639.
}


\appendix

\section{Invertible conformal transformation with the curvature tensor}\label{AppA}

In the main text, we focused on invertible transformations where derivatives of the metric appear only through covariant derivatives of the scalar field.
Interestingly, one can also construct invertible transformations containing the curvature tensors in the transformation law for the metric.
For instance, the following conformal transformation is invertible in general:
    \be
    \bar{g}_\mn=\Omega(\mC)g_\mn, \qquad
    \mC\coloneqq \mW^{\alpha\beta\gamma\delta}\mW_{\alpha\beta\gamma\delta},
    \ee
where $\Omega$ is an arbitrary nonvanishing function and $\mW^\mu{}_{\nu\lambda\sigma}$ denotes the Weyl tensor.
The point is that the quantity~$\mC$ is transformed under the conformal transformation as
    \be
    \bar{\mC}=\Omega^{-2}\mC, \label{Weylsq_conformal}
    \ee
which allows us to express $\bar{\mC}$ as a function of $\mC$ (at least locally) unless $\Omega\propto \mC^{1/2}$.
Hence, the inverse transformation is given explicitly by
    \be
    g_\mn=\bar{\Omega}(\bar{\mC})\bar{g}_\mn, \qquad
    \bar{\Omega}(\bar{\mC})\coloneqq \Omega^{-1}(\mC).
    \ee
Similarly, with the use of conformally (or disformally) invariant quantities, one can construct invertible conformal (or disformal) transformations containing higher derivatives of the scalar field~\cite{Domenech:2019syf}.

Note that $\mC$ above can be replaced by the Chern-Simons (or Pontryagin) term~\cite{Jackiw:2003pm,Grumiller:2007rv}
    \be
    \mathcal{P}\coloneqq \fr{1}{2}\vae^{\alpha\beta\gamma\delta}R^{\mu\nu}{}_{\alpha\beta}R_{\mu\nu\gamma\delta}
    = \fr{1}{2}\vae^{\alpha\beta\gamma\delta}\mW^{\mu\nu}{}_{\alpha\beta}\mW_{\mu\nu\gamma\delta},
    \ee
with $\vae^{\alpha\beta\gamma\delta}$ being the totally antisymmetric tensor.
This is because the quantity~$\mathcal{P}$ is also transformed as \eqref{Weylsq_conformal} under a conformal transformation.
Moreover, one can consider a general function~$\Omega(\mC,\mathcal{P})$.
In this case, the transformation is invertible so long as
	\be
	\left|\fr{\pa(\bar{\mC},\bar{\mathcal{P}})}{\pa(\mC,\mathcal{P})}\right|
	\propto \Omega-2\mC\fr{\pa\Omega}{\pa\mC}-2\mathcal{P}\fr{\pa\Omega}{\pa\mathcal{P}}\ne 0.
	\ee
By use of this novel class of invertible transformations, one can generate degenerate higher-derivative metric theories from general relativity.
(For an earlier attempt to construct degenerate higher-derivative metric theories, see \cite{Crisostomi:2017ugk}.)

\section{Consistent matter coupling in two-DOF theories}\label{AppB}

In \S\ref{sec:coupling}, we studied conditions under which a matter field can be consistently coupled to GDH theories.
We stress that the conditions obtained there apply only to generic cases where the scalar field is dynamical, and a separate analysis is required for special cases where the scalar field is nondynamical (e.g., the cuscuton~\cite{Afshordi:2006ad} or its extension~\cite{Iyonaga:2018vnu,Iyonaga:2020bmm}).
This kind of theories is sometimes called {\it minimally modified gravity} since there exist just two propagating DOFs associated with the dynamical metric as in general relativity.
In such theories, the gravitational sector of the action satisfies an additional degeneracy condition compared to the generic case, and hence the matter sector should also respect it.
To the best of our knowledge, the condition for consistent matter coupling has not been specified even for two-DOF theories within disformal Horndeski theories, i.e., those related to Horndeski theories via invertible conventional disformal transformation.
In this appendix, we clarify the consistency condition at the level of linear cosmological perturbations.

As we did in \S\ref{ssec:coupling_linear}, we consider cosmological scalar perturbations with a k-essence (matter) scalar field described by the action~\eqref{matter_action} and work in the frame where the gravitational action is given by the Horndeski one.
Note that we will impose a condition coming from the two-DOF nature of the gravitational sector when it is necessary.
For the conventional disformal coupling where $\bar{g}_\mn=F_0(\phi,X)g_\mn+F_1(\phi,X)\phi_\mu\phi_\nu$, the quadratic Lagrangian for the matter sector can be written in the form
    \be
    \mL^{(2)}_{\rm m}=
    Na^3\mJ\bra{\fr{1}{2}b_1\fr{\dot{\delta\sigma}^2}{N^2}+b_2\alpha\fr{\dot{\delta\sigma}}{N}+b_3\zeta\fr{\dot{\delta\sigma}}{N}+\fr{1}{2}b_4\alpha^2+b_5\delta\sigma\fr{\pa^2\chi}{a}+\fr{1}{2}b_6\delta\sigma\fr{\pa^2\delta\sigma}{a^2}},
    \ee
where $\mJ=F_0^{3/2}f^{1/2}$ with $\mFc\coloneqq F_0+XF_1$ and the coefficients are given by
    \be
    \begin{split}
    &b_1=-\fr{2}{\mFc}\bra{P'+2\kinmat P''}, \qquad
    b_2=\fr{2\dot{\sigma}}{N}\brb{\fr{1}{F_0^3}\bra{\fr{XF_0^3}{\mFc}}_XP'+2\bra{\fr{X}{\mFc}}_X\kinmat P''}, \\
    &b_3=-\fr{6}{\mFc}\fr{\dot{\sigma}}{N}P', \qquad
    b_4=-\fr{X\mJ_{QQ}}{\mJ}P
    -\fr{\mFc}{\mJ^2}\brb{\mJ^2\bra{\fr{X}{\mFc}}_Q}_Q\kinmat P'
    +4\brb{\mFc\bra{\fr{X}{\mFc}}_X}^2\kinmat^2P'', \\
    &b_5=-\fr{2}{\mFc}\fr{\dot{\sigma}}{N}P', \qquad
    b_6=-\fr{2}{F_0}P'.
    \end{split}
    \ee
Here, $\kinmat=-\dot{\sigma}^2/(fN^2)$ is the kinetic term of the matter scalar field which is coupled to the disformal metric, and a subscript~$Q$ denotes the differentiation with respect to $Q\coloneqq \sqrt{-X}$, e.g.,
    \be
    \mJ_Q=-2\sqrt{-X}\mJ_X, \qquad
    \mJ_{QQ}=-2\bra{\mJ_X+2X\mJ_{XX}}.
    \ee
Note that the quantity~$\mFc$ is related to $\mF$ defined in \eqref{mF}.
Indeed, for conventional disformal transformations where $F_2=F_3=0$, we have $\mF=F_0 \mFc$.
Going to the Fourier space, the total quadratic Lagrangian~$\mL^{(2)}=\mL^{(2)}_{\rm H}+\mL^{(2)}_{\rm m}$ can be written in the form
    \be
    \mL^{(2)}=\sum_{A,B=1}^{4}\bra{\fr{1}{2}\mK_{AB}\dot{v}^A\dot{v}^B+\mM_{AB}\dot{v}^Av^B-\fr{1}{2}\mW_{AB}v^Av^B},
    \ee
up to total derivative, where $v^A \coloneqq(\alpha,\chi,\zeta,\delta\sigma)$ and the coefficient matrices depend on the wavenumber.
[See Eq.~\eqref{qLag_H} for the expression of $\mL^{(2)}_{\rm H}$.]
Note that the matrices~$\mK$ and $\mW$ are symmetric and $\mM$ is antisymmetric.
Then, the dispersion relation for scalar perturbations can be computed by
    \be
    \det\brb{\omega^2\mK_{AB}+i\omega\bra{2\mM_{AB}+\dot{\mK}_{AB}}-\bra{\mW_{AB}+\dot{\mM}_{AB}}}=0, \label{disp_rel_conventional}
    \ee
which yields a quartic algebraic equation for $\omega$ in general.
Namely, there exist two propagating DOFs, which can be identified with two dynamical scalar fields~$\phi$ and $\sigma$.
This implies that there is no unwanted extra DOF for generic Horndeski theories in the case of conventional disformal coupling, which was already pointed out in \cite{Deffayet:2020ypa}.

On the other hand, for scalar-tensor theories where the scalar field~$\phi$ is nondynamical, the presence of $\omega^4$~term in \eqref{disp_rel_conventional} implies the inconsistency of matter coupling.
The coefficient of the $\omega^4$~term in the dispersion relation is proportional to the following quantity:
    \be
    (b_2^2-b_1b_4)\mJ\mG_T-2b_1(3\Theta^2+\mG_T\Sigma).
    \ee
In order for the matter coupling to be consistent, we require that this coefficient should vanish.
For the (extended) cuscutons~\cite{Afshordi:2006ad,Iyonaga:2018vnu,Iyonaga:2020bmm} where $\mG_S\propto3\Theta^2+\mG_T\Sigma=0$ is satisfied,\footnote{Note that the quantity~$3\Theta^2+\mG_T\Sigma$ is a polynomial of the Hubble parameter whose coefficients are written in terms of the functions that characterize the gravitational action. The extended cuscutons were constructed by requiring that each of these coefficients should vanish independently without using the background equations of motion.} this requirement yields
    \be
    b_2^2-b_1b_4
    =\fr{24X^2F_{0X}^2}{F_0^5\mJ^2}\mB_1[P]
    +\fr{12X\brb{F_{0X}\bra{\mJ+4X\mJ_X}+2X\mJ F_{0XX}}}{F_0^4\mJ^3}\mB_2[P]
    -\fr{2X\mJ_{QQ}}{F_0^3\mJ^3}\mB_3[P]=0. \label{omega4}
    \ee
Here, $\mB$'s are linearly independent functionals of $P(\kinmat)$ given by
	\be
	\begin{split}
	&\mB_1=2PP'-2\kinmat P'^2+4\kinmat P P''-\kinmat^2 P' P'', \qquad
	\mB_2=(P-\kinmat P')(P'+2\kinmat P''), \\
	&\mB_3=(P-2\kinmat P')(P'+2\kinmat P'').
	\end{split}
	\ee
Since \eqref{omega4} should be satisfied for any function~$P$, we have
    \be
    F_{0X}=F_{0XX}=\mJ_{QQ}=0, \label{consistency}
    \ee
which poses constraints on the functions~$F_0$ and $F_1$ that define the disformal transformation.
Although the above conditions are required only on the background trajectory in the configuration space~$(\phi,X)$, one could impose them over the entire configuration space.
Then, the functional form of $F_0$ and $\mJ$ is fixed as
	\be
	F_0=F_0(\phi), \qquad
	\mJ=\mJ_0(\phi)+\mJ_1(\phi)\sqrt{-X}, \label{consistency_explicit}
	\ee
where $F_0$, $\mJ_0$, and $\mJ_1$ are arbitrary functions of $\phi$ such that $F_0\ne 0$ and $(\mJ_0,\mJ_1)\ne (0,0)$.\footnote{The authors of \cite{Iyonaga:2021yfv} studied a two-DOF theory that is obtained from the extended cuscutons via disformal transformation satisfying \eqref{consistency_explicit}. They found that the theory can be consistently coupled with a canonical scalar field, which is consistent with our result.}
The functional form of $F_1$ is given by
	\be
	F_1=\fr{\brb{\mJ_0(\phi)+\mJ_1(\phi)\sqrt{-X}}^2-F_0(\phi)^4}{F_0(\phi)^3 X}.
	\ee
Note that the choice of $\mJ$ in \eqref{consistency_explicit} implies that
	\be
	\sqrt{-\bar{g}}=\sqrt{-g}\,\mJ
	=\sqrt{\gamma}\brb{\mJ_0(\phi)N+\dot{\phi}\mJ_1(\phi)},
	\ee
under the unitary gauge where $\sqrt{-X}=\dot{\phi}/N$.
Namely, $\sqrt{-\bar{g}}$ depends on $N$ only up to the linear order.
In principle, it would be possible to generalize the above discussion to generalized disformal transformations, which is left for future study.

\section{Canonical form of singular transformations}\label{AppC}

In the main text, we focused on invertible disformal transformations satisfying the condition~\eqref{inv_cond}.
On the other hand, singular (or noninvertible) transformations have been studied in the context of mimetic gravity~\cite{Chamseddine:2013kea} (see also \cite{Sebastiani:2016ras} for a review).
In this appendix, we focus on a singular subclass of generalized disformal transformations to show that such transformations can be brought to a simpler form by use of an appropriate invertible generalized disformal transformation.

Let us first consider the case of conventional disformal transformations of the form
    \be
    \bar{g}_\mn[g,\phi] = F_0(\phi,X) g_\mn + F_1(\phi,X)\phi_\mu\phi_\nu,
    \ee
for which $\bar{X}=\bar{g}^\mn\phi_\mu\phi_\nu=X/(F_0+X F_1)$.
Singular transformations of our interest are characterized by the following conditions:
	\be
	F_0\ne 0, \qquad
	F_0+X F_1\ne 0, \qquad
	\bar{X}_X(\phi,X)=0. \label{noninv_cond_conv}
	\ee
The simplest example of such singular transformations would be the following conformal one:
	\be
	\bar{g}_\mn[g,\phi]=\fr{X}{c(\phi)}g_\mn, \label{canonical_noninv_conv}
	\ee
with $c(\phi)$ being an arbitrary (nonvanishing) function of $\phi$.
For this transformation, the barred inverse metric is given by $\bar{g}^\mn=c(\phi)X^{-1}g^\mn$ and hence $\bar{X}=\bar{g}^\mn\phi_\mu\phi_\nu=c(\phi)$.
The noninvertibility of this particular transformation can be understood by the conformal invariance of the right-hand side of the transformation law.
Indeed, under any conformal transformation~$g_\mn\to \Omega g_\mn$, the right-hand side of \eqref{canonical_noninv_conv} remains unchanged.
Due to the conformal invariance, the transformation law cannot be solved uniquely for $g_\mn$.
It was shown in \cite{Takahashi:2017pje} that any singular disformal transformation satisfying \eqref{noninv_cond_conv} can be brought to the simplest form~\eqref{canonical_noninv_conv} by use of an appropriate invertible (conventional) disformal transformation.
In this sense, the noninvertible conformal transformation~\eqref{canonical_noninv_conv} can be regarded as the canonical form of singular conventional disformal transformations.

In what follows, we discuss a singular subclass of the generalized disformal transformation~\eqref{disformal2} and prove an extension of the above statement.
In this case, the invertibility conditions were given in \eqref{inv_cond}, which we recapitulate here for convenience:
	\be
	F_0\ne 0, \qquad
	\mF\ne 0, \qquad
	\bar{X}_Y=\bar{X}_Z=0, \qquad
	\bar{X}_X\ne 0, \qquad
	\left|\fr{\pa(\bar{Y},\bar{Z})}{\pa(Y,Z)}\right|\ne 0.
	\label{inv_cond_AppC}
	\ee
The first two conditions are responsible for the existence of the inverse metric~$\bar{g}^\mn$, while the third one is necessary for the closedness under the functional composition of two disformal transformations (see \S\ref{sec:inv}).
In what follows, we study generalized disformal transformations that do not satisfy the fourth condition in \eqref{inv_cond_AppC}, i.e., we focus on transformations characterized by the following conditions:
	\be
	F_0\ne 0, \qquad
	\mF\ne 0, \qquad
	\bar{X}=c(\phi),
	\label{generalized_noninv_cond}
	\ee
with $c(\phi)$ being a nonvanishing function of $\phi$.
This is a generalization of the singular subclass of conventional disformal transformations satisfying \eqref{noninv_cond_conv}.
We shall argue that the singular conformal transformation~\eqref{canonical_noninv_conv} serves also as the canonical form of singular generalized disformal transformations satisfying \eqref{generalized_noninv_cond}.
Note in passing that there also exists another type of singular transformations satisfying $\bar{X}_X\ne 0$ but not the last condition in \eqref{inv_cond_AppC}.
A specific example is given by 
    \be
    \bar{g}_\mn[g,\phi]
    =\fr{hZ}{Y^2-\bar{X}_0hZ^2}\brb{Wg_\mn+\bra{X-\bar{X}_0h}X_\mu X_\nu}, \qquad
    h=h\bra{\fr{Y}{Z}}, \qquad
    \bar{X}_0=\bar{X}_0(\phi,X), \label{noninv_example}
    \ee
with $h\ne 0$ and $\bar{X}_{0X}\ne 0$.
These transformations may also provide an interesting class of singular transformations, which we leave for future study.

Let us now consider the following two transformations:
    \be
    \begin{split}
    \bar{g}_\mn[g,\phi] &= F_0 g_\mn + F_1 \phi_\mu\phi_\nu
    + 2F_2 \phi_{(\mu}X_{\nu)} + F_3 X_\mu X_\nu, \quad \text{(singular)} \\
    \hat{g}_\mn[g,\phi] &= f_0 g_\mn + f_1 \phi_\mu\phi_\nu
    + 2f_2 \phi_{(\mu}X_{\nu)} + f_3 X_\mu X_\nu, \quad \text{(invertible)}
    \end{split}
    \ee
where $F_i$'s and $f_i$'s are functions of $(\phi,X,Y,Z)$.
Here, we assume that $\bar{g}_\mn[g,\phi]$ is singular 
and $\hat{g}_\mn[g,\phi]$ is invertible.
We shall prove that, for any singular transformation~$\bar{g}_\mn[g,\phi]$ with the coefficient functions~$F_i$ satisfying \eqref{generalized_noninv_cond}, there exists an invertible transformation~$\hat{g}_\mn[g,\phi]$ such that the functional composition~$(\bar{g}\circ\hat{g})_\mn[g,\phi]$ coincides with the singular conformal transformation~\eqref{canonical_noninv_conv}.
As in the main text, we require the existence of the inverse metrics~$\bar{g}^\mn$ and $\hat{g}^\mn$, which are used to define the barred/hatted counterparts of the kinetic term of the scalar field,
	\begin{align}
	\bar{X}[g,\phi]&\coloneqq \bar{g}^\mn[g,\phi]\phi_\mu\phi_\nu
	=\fr{XF_0-WF_3}{F_0^2+F_0(XF_1+2YF_2+ZF_3)+W(F_2^2-F_1F_3)}, \label{Xbar_appC} \\
	\hat{X}[g,\phi]&\coloneqq \hat{g}^\mn[g,\phi]\phi_\mu\phi_\nu
	=\fr{Xf_0-Wf_3}{f_0^2+f_0(Xf_1+2Yf_2+Zf_3)+W(f_2^2-f_1f_3)}. \label{Xhat_appC}
	\end{align}
As explained in \S\ref{sec:inv}, $\bar{X}[g,\phi]$ and $\hat{X}[g,\phi]$ must be functions of $(\phi,X)$ to guarantee the closedness under the functional composition.
Also, as mentioned earlier, we assume that the singular transformation has $\bar{X}=c(\phi)$, with $c$ being a nonvanishing function of $\phi$ [see Eq.~\eqref{generalized_noninv_cond}].

Under this setup, the functional composition of the two disformal transformations is given by
    \begin{align}
    (\bar{g}\circ\hat{g})_\mn[g,\phi]
    &=F_0 \hat{g}_\mn + F_1 \phi_\mu\phi_\nu
    + 2F_2 \phi_{(\mu}\hat{X}_{\nu)} + F_3 \hat{X}_\mu \hat{X}_\nu \nonumber \\
    &=F_0f_0g_\mn + (F_0f_1+F_1+2\hat{X}_\phi F_2+\hat{X}_\phi^2F_3) \phi_\mu\phi_\nu \nonumber \\
    &\quad + 2(F_0f_2+\hat{X}_XF_2+\hat{X}_\phi\hat{X}_XF_3) \phi_{(\mu}X_{\nu)} + (F_0f_3+\hat{X}_X^2F_3) X_\mu X_\nu. \label{func_comp_singular}
    \end{align}
The functions~$F_i$ are now evaluated at $(\phi,\hat{X},\hat{Y},\hat{Z})$, with
	\be
	\begin{split}
	\hat{Y}&\coloneqq \hat{g}^\mn\phi_\mu \hat{X}_\nu
	=\fr{\hat{X}_X(Y f_0 + W f_2)}{f_0^2+f_0(Xf_1+2Yf_2+Zf_3)+W(f_2^2-f_1f_3)}+\hat{X}_\phi\hat{X}, \\
	\hat{Z}&\coloneqq \hat{g}^\mn\hat{X}_\mu \hat{X}_\nu
	=\fr{\hat{X}_X^2(Zf_0-Wf_1)}{f_0^2+f_0(Xf_1+2Yf_2+Zf_3)+W(f_2^2-f_1f_3)}+2\hat{X}_\phi\hat{Y}-\hat{X}_\phi^2\hat{X}.
	\end{split} \label{YZhat}
	\ee
It is also useful to introduce the following quantity:
    \be
	\hat{W}\coloneqq \hat{Y}^2-\hat{X}\hat{Z}
	=\fr{\hat{X}_X^2W}{f_0^2+f_0(Xf_1+2Yf_2+Zf_3)+W(f_2^2-f_1f_3)}. \label{What}
	\ee
One could also consider the functional composition of the reversed order, i.e., $\hat{g}\circ\bar{g}$. However, in this case, the functional composition cannot be brought to the conformal form in general.
In any case, we are interested in the functional composition~\eqref{func_comp_singular} for the reason explained below.

Let us clarify conditions under which the functional composition~\eqref{func_comp_singular} coincides with the singular conformal transformation~\eqref{canonical_noninv_conv}.
Requiring that the disformal factors in \eqref{func_comp_singular} vanish, the functions~$f_1$, $f_2$, and $f_3$ satisfy
    \be
    f_1=-\fr{F_1+2\hat{X}_\phi F_2+\hat{X}_\phi^2F_3}{F_0}, \qquad
    f_2=-\fr{\hat{X}_XF_2+\hat{X}_\phi\hat{X}_XF_3}{F_0}, \qquad
    f_3=-\fr{\hat{X}_X^2F_3}{F_0}. \label{f2F}
    \ee
The functional form of $f_0$ is fixed by the condition [see Eq.~\eqref{Xbar_appC}]
	\be
	\bar{X}[\hat{g},\phi]=\fr{\hat{X}F_0-\hat{W}F_3}{F_0^2+F_0(\hat{X}F_1+2\hat{Y}F_2+\hat{Z}F_3)+\hat{W}(F_2^2-F_1F_3)}=c(\phi).
	\ee
Plugging \eqref{Xhat_appC} and \eqref{YZhat}--\eqref{f2F} into this condition, we obtain 
\be f_0=\fr{X}{c(\phi)F_0}. \label{f0F} \ee 
Hence, \eqref{func_comp_singular} reads
    \be
    (\bar{g}\circ\hat{g})_\mn[g,\phi]=\fr{X}{c(\phi)}g_\mn. \label{canonical_noninv}
    \ee
Note that the functional form of $\hat{X}(\phi,X)$ remains unfixed and it can be an arbitrary function so long as $\hat{X}_X\ne 0$ (as well as the other invertibility conditions for $\hat{g}_\mn[g,\phi]$ are satisfied).

In summary, for any given singular generalized disformal transformation~$\bar{g}_\mn[g,\phi]$ with the coefficient functions~$F_i$ satisfying \eqref{generalized_noninv_cond}, one can choose an invertible transformation~$\hat{g}_\mn[g,\phi]$ such that the functional composition~$(\bar{g}\circ\hat{g})_\mn[g,\phi]$ is the singular conformal transformation.
The coefficient functions~$f_i$ that characterize the invertible transformation can be fixed up to an arbitrary choice of the function~$\hat{X}(\phi,X)$ with $\hat{X}_X\ne 0$.
Indeed, $f_0$ is given by \eqref{f0F}, and then \eqref{f2F} provides a system of algebraic equations that fixes $f_1$, $f_2$, and $f_3$.\footnote{In general, the algebraic equations have a nonlinear dependence on $f_i$'s, and hence the existence of a solution may be nontrivial.
One may choose a simplest $\hat{X}(\phi,X)$ so that the system of algebraic equations has a (global) solution.}
Note that, as mentioned above, $F_i$'s in \eqref{f2F} are evaluated at $(\phi,\hat{X},\hat{Y},\hat{Z})$, which can be rewritten as functions of $(\phi,X,Y,Z)$ by use of \eqref{YZhat}.
Finally, one should check whether the so-obtained transformation~$\hat{g}_\mn[g,\phi]$ is invertible.
If not, one may try another choice of $\hat{X}(\phi,X)$ to make the transformation invertible.

One may find the above simple result surprising, but this is actually expected.
Indeed, since $\bar{g}_\mn[g,\phi]$ is singular, the functional composition~$(\bar{g}\circ\hat{g})_\mn[g,\phi]$ must be singular as well.
Also, we fixed $\hat{g}_\mn[g,\phi]$ so that $(\bar{g}\circ\hat{g})_\mn[g,\phi]\propto g_\mn$.
Then, $(\bar{g}\circ\hat{g})_\mn[g,\phi]$ is a singular conformal transformation, which must be of the form~\eqref{canonical_noninv_conv}.
The above discussion shows that \eqref{canonical_noninv_conv} can be regarded as the canonical form of singular generalized disformal transformations satisfying \eqref{generalized_noninv_cond}, which is a natural extension of the result of \cite{Takahashi:2017pje}.

Let us demonstrate the above result for the following singular generalized disformal transformation:
    \be
    \bar{g}_\mn[g,\phi]=g_\mn+F_3(\phi,X,Y,Z)X_\mu X_\nu, \qquad
    F_3(\phi,X,Y,Z)\coloneqq \fr{X-c(\phi)}{Y^2-XZ+c(\phi)Z}, \qquad
    c(\phi)\ne 0,
    \label{singular_example}
    \ee
for which $\bar{X}=c(\phi)$.\footnote{Actually, with $\bar{X}=c(\phi)$, $F_0=1$, and $F_1=F_2=0$, the functional form of $F_3$ is fixed to the one in \eqref{singular_example} by use of \eqref{F3}.}
In this case, we set $F_0=1$ and $F_1=F_2=0$ in the above discussion.
As mentioned above, one can choose an arbitrary $\hat{X}(\phi,X)$ as long as $\hat{X}_X\ne 0$, so we simply set $\hat{X}_\phi=0$ and hence $\hat{X}=\hat{X}(X)$.
Then, \eqref{f2F} and \eqref{f0F} read
    \be
    f_0=\fr{X}{c}, \qquad 
    f_1=f_2=0, \qquad
    f_3=-\hat{X}_X^2 F_3. \label{f2F_example}
    \ee
Note that $F_3$ in \eqref{f2F_example} is still evaluated at $(\phi,\hat{X},\hat{Y},\hat{Z})$, which we need to rewrite in terms of unbarred quantities.
From \eqref{YZhat}, we obtain 
    \be
    \hat{Y}=\fr{c\hat{X}_X Y}{X+cZf_3}, \qquad
    \hat{Z}=\fr{c\hat{X}_X^2 Z}{X+cZf_3},
    \ee
and hence
    \be
    F_3(\phi,\hat{X},\hat{Y},\hat{Z})=\fr{(\hat{X}-c)(X+cZf_3)^2}{c\hat{X}^2\brb{cY^2-Z(\hat{X}-c)(X+cZf_3)}}. \label{F3_example}
    \ee
From \eqref{f2F_example} and \eqref{F3_example}, the functional form of $f_3$ is determined as
    \be
    f_3(\phi,X,Y,Z)=-\fr{X^2(\hat{X}-c)}{c^2(Y^2-XZ)+cX\hat{X}Z}.
    \ee
Therefore, one obtains the canonical singular transformation~\eqref{canonical_noninv} by the functional composition of the singular transformation~\eqref{singular_example} and the following transformation:
    \be
    \hat{g}_\mn[g,\phi] = \fr{X}{c}g_\mn -\fr{X^2(\hat{X}-c)}{c^2(Y^2-XZ)+cX\hat{X}Z} X_\mu X_\nu, \label{ghat_example}
    \ee
with $\hat{X}$ being an arbitrary nonvanishing function of $X$.
Also, it is straightforward to check that the transformation~\eqref{ghat_example} satisfies the invertibility condition~\eqref{inv_cond_AppC}.

Finally, let us explain the implication of the above result when the disformal transformations act on scalar-tensor theories.
Suppose that we start from some seed scalar-tensor theory described by the action~$S[g,\phi]$ and perform sequential replacements~$g\to \bar{g}[g,\phi]$ and then $g\to \hat{g}[g,\phi]$.
The first replacement yields $S'[g,\phi]\coloneqq S[\bar{g},\phi]$, and then the second one yields $S''[g,\phi]\coloneqq S'[\hat{g},\phi]$.
Note that the action~$S''[g,\phi]$ can be directly obtained from the seed action~$S[g,\phi]$ by the singular conformal transformation~$g\to \bar{g}\circ\hat{g}[g,\phi]$, and hence it is nothing but the action of conventional mimetic gravity models.
On the other hand, the theory described by the action~$S'[g,\phi]$ is a generalization of known mimetic gravity models as it is obtained by performing the singular generalized disformal transformation~$g\to \bar{g}[g,\phi]$.
Our result implies that the generalized mimetic theory~$S'[g,\phi]$ is related to the conventional mimetic theory~$S''[g,\phi]$ via invertible generalized disformal transformation.
This is why we focused on the functional composition~$\bar{g}\circ\hat{g}$, not $\hat{g}\circ\bar{g}$.
(See Fig.~\ref{fig2} below.)
Of course, the two theories~$S'[g,\phi]$ and $S''[g,\phi]$ are distinguished with each other when matter fields are taken into account.
Even so, when one studies the generalized mimetic theory~$S'[g,\phi]$, our result can be used to move to the frame where the gravitational action is given by that of the conventional mimetic theory, which could simplify the analysis.

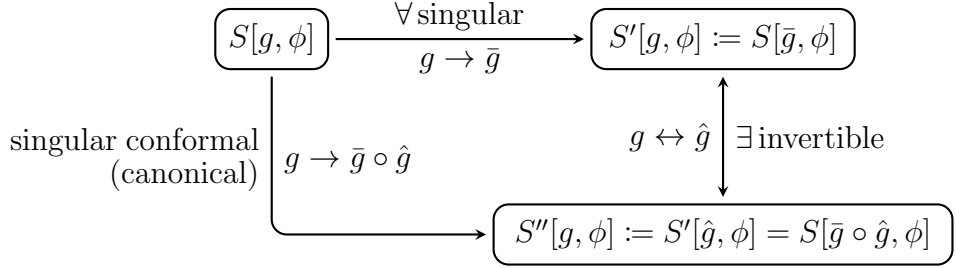
\begin{figure}[H]
\centering
\begin{tikzpicture}
	\draw[thick, rounded corners=0.2cm] (0,5)--(1.5,5)--(1.5,4.2)--(0,4.2)--cycle;
	\node[anchor=center] at (0.75,4.6) {\large{$S[g,\phi]$}};
	\draw[thick, rounded corners=0.2cm] (5,5)--(8.5,5)--(8.5,4.2)--(5,4.2)--cycle;
	\node[anchor=center] at (6.75,4.6) {\large{$S'[g,\phi]\coloneqq S[\bar{g},\phi]$}};
	\draw[thick, rounded corners=0.2cm] (3.7,2.4)--(9.8,2.4)--(9.8,1.6)--(3.7,1.6)--cycle;
	\node[anchor=center] at (6.75,2) {\large{$S''[g,\phi]\coloneqq S'[\hat{g},\phi]=S[\bar{g}\circ\hat{g},\phi]$}};
	\draw[->, >=stealth, thick] (1.6,4.6)--(4.9,4.6);
	\node[anchor=center] at (3.25,4.3) {\large{$g\to \bar{g}$}};
	\node[anchor=center] at (3.25,4.9) {\large{$\forall$\,singular}};
	\draw[<->, >=stealth, thick] (6.75,4.1)--(6.75,2.5);
	\node[anchor=west] at (6.8,3.3) {\large{$\exists$\,invertible}};
	\node[anchor=east] at (6.7,3.3) {\large{$g\leftrightarrow \hat{g}$}};
	\draw[->, >=stealth, thick, rounded corners=0.2cm] (0.75,4.1)--(0.75,2)--(3.6,2);
	\node[anchor=west] at (0.8,3) {\large{$g\to \bar{g}\circ\hat{g}$}};
	\node[rectangle, text width=3.35cm, minimum height=1cm, align=right, anchor=east] at (0.7,3) {\large{singular conformal\\(canonical)}};
\end{tikzpicture}
\caption{Functional composition of singular and invertible disformal transformations acting on scalar-tensor theories.
For any given singular generalized disformal transformation~$g\to \bar{g}[g,\phi]$ satisfying \eqref{generalized_noninv_cond}, there exists an invertible transformation~$g\to \hat{g}[g,\phi]$ such that $g\to \bar{g}\circ\hat{g}[g,\phi]$ is the canonical singular transformation~\eqref{canonical_noninv}.}
\label{fig2}
\end{figure}


\bibliographystyle{mybibstyle}
\bibliography{bib}

\end{document}